\documentclass[a4paper,12pt]{spieman}  
\usepackage{amsmath,amsfonts,amssymb}
\usepackage{graphicx}
\usepackage{setspace}
\usepackage{tocloft}
\usepackage{lineno}

\usepackage{verbatim}
\newcommand{\xsp}{XSPECT\,}
\graphicspath{{Figure_files_one_per_page/}}
\usepackage{subcaption}	
\usepackage[usenames,dvipsnames]{xcolor}
\usepackage{threeparttable}
\usepackage{multirow}
\usepackage{enumitem}
\usepackage{pdfcolfoot}  

\title{XSPECT on-board XPoSat: Calibration and First Results}

\author[a,*]{Rwitika Chatterjee}
\author[a]{Koushal Vadodariya}
\author[a]{Radhakrishna Vatedka}
\author[a]{Vivek Kumar Agrawal}
\author[a]{Anurag Tyagi}
\author[a]{Kiran M Jayasurya}
\author[a]{Shyam Prakash V. P.}
\author[a]{Ramadevi M C}
\author[a]{Vaishali Sharan}
\affil[a]{Space Astronomy Group, U R Rao Satellite Center, ISITE Campus, Outer Ring Road, Karthik Nagar, Bengaluru, Karnataka 560037, India}

\cftpagenumbersoff{figure}
\cftpagenumbersoff{table}

\begin{document} 
\maketitle

\begin{abstract}
XPoSat is India's first X-ray spectro-polarimetry mission, consisting of two co-aligned instruments, a polarimeter (POLIX) and a spectrometer (XSPECT), to study the X-ray emission from celestial sources. Since polarimetry is a photon-hungry technique, the mission is designed to observe sources for long integration times ($\sim$ few days to weeks). This provides an unique opportunity, enabling XSPECT to carry out long-term monitoring of sources, and study their spectro-temporal evolution. To ensure that the instrument is able to fulfill its scientific objectives, it was extensively calibrated on-ground. Post launch, these calibrations were validated using on-board observations. Additionally, some aspects of the instrument such as alignment and effective area were also derived and fine-tuned from in-flight data. In this paper, we describe the calibration of \xsp instrument in detail, including some initial results derived from its data to establish its capabilities.
\end{abstract}

\keywords{X-ray spectroscopy, X-ray timing, XPoSat, XSPECT, Calibration}

{\noindent \footnotesize\textbf{*}Rwitika Chatterjee,  \linkable{rwitika@ursc.gov.in} }

\begin{spacing}{1.2}   

\section{Introduction}
\label{sec:intro}  
XPoSat is India's second dedicated astronomy mission after AstroSat [\citenum{singh2014}]. The primary payload, Polarimeter Instrument in X-rays (POLIX, [\citenum{paul2016}]), is the first ever medium-energy ($8-30$~keV) polarimeter to be flown. Additionally, XPoSat also includes a soft X-ray spectrometer \xsp (X-ray SPECtroscopy \& Timing) sensitive to X-rays in the energy band $0.8-15$~keV [\citenum{rkrish2025}]. The primary targets of XSPECT include neutron stars and black hole sources (compact objects), typically in binary systems, accreting matter from a `normal' companion star. The main scientific areas that \xsp aims to address are: (1) understanding the nature, origin and variability of the soft excess X-ray pulsars, (2) investigating the evolution of spin period and pulse profile, (3) measuring the spin and mass of black holes through continuum and iron line profile fitting, (4) studying evolution of X-ray spectrum in neutron star low-mass X-ray binaries and nature of soft thermal component, and (5) study of low frequency quasi-periodic oscillations (QPOs) in X-ray binaries in soft X-ray band. Taking advantage of the long duration observations required by POLIX to measure polarization, XSPECT can carry out long-term monitoring of spectral state changes in continuum emission, changes in their line flux and profile, and simultaneous temporal monitoring of soft X-ray emission. 

XPoSat was launched on $1^{\mathrm{st}}$~January,~2024 by PSLV-C58, from Sriharikota, India, and was placed into an equatorial low-Earth orbit with 650~km altitude and $6^{\circ}$ inclination. XPoSat orbits the Earth $\sim14-15$~times per Earth-day, with each orbit lasting for $\sim100$~minutes. The science instruments are mounted facing a direction opposite to that of the solar panels. The source observations are carried out only during the eclipse (`night side') of the orbit, during which period the satellite is also slowly spun about its axis ($\sim 1.2^{\circ}$~s$^{-1}$) to eliminate the effect of systematics in polarization measurements by POLIX. On the day side, the solar panels are aligned normal to the Sun for optimal power generation, when \xsp mainly sees the solar-illuminated Earth in its field of view (FOV). The eclipse passes of $1-2$ orbits per day over the ground station is reserved for the download of payload as well as auxiliary data. Being in an equatorial orbit, the satellite also passes through the South Atlantic Anomaly (SAA) every orbit, which is a region of increased flux of energetic particles. \xsp is switched OFF over the duration of SAA passage to protect the detectors from high dosage of ionizing radiation which may lead to accelerated degradation. Apart from this period, \xsp is always ON and recording data. The day-averaged duty cycle of source observations is $\sim 20\%$. Further details about XPoSat mission design, planning and operations can be found in [\citenum{saini2025}].

On January 2, 2024, \xsp was switched on, which also marked the commencement of the performance verification (PV) phase of the instrument. To ensure that \xsp is capable of meeting its scientific objectives, the instrument was extensively calibrated on-ground, which was further validated and fine-tuned using on-board observations. This paper describes the on-ground as well as initial on-board calibration (during the PV phase) of the \xsp payload, and also demonstrates its capability using the initial observations and first results obtained from it.

\subsection{XSPECT Instrument}
\label{subsec:xsp}

The detectors used in \xsp are second generation Swept Charge Devices (SCDs), viz. CCD236, developed by e2V Technologies [\citenum{lowe2001, holland2008}]. SCDs are similar to the familiar Charge Coupled Devices (CCDs), but employ a different readout scheme (see Section~\ref{sec:sinevt}), which, although, results in a loss of positional information of the events, but are much faster to read out, giving the advantage of pile-up free observation of bright sources. By simulating the charge flow and readout of the SCD, the pile-up fraction is estimated to be $< 1\%$ for source intensity of upto 60 Crab. Hence, \xsp provides an unique platform for the study of bright X-ray sources as well as future bright transients. SCDs have been previously flown in C1XS [\citenum{grande2009}] on-board Chandrayaan-1, CLASS [\citenum{rkrish2020}] on-board Chandrayaan-2 and LE telescopes [\citenum{chen2020}] on-board Insight-HXMT.

\begin{figure*}
\centering
\includegraphics[scale=0.6, trim=0cm 2cm 0cm 2cm, clip=true]{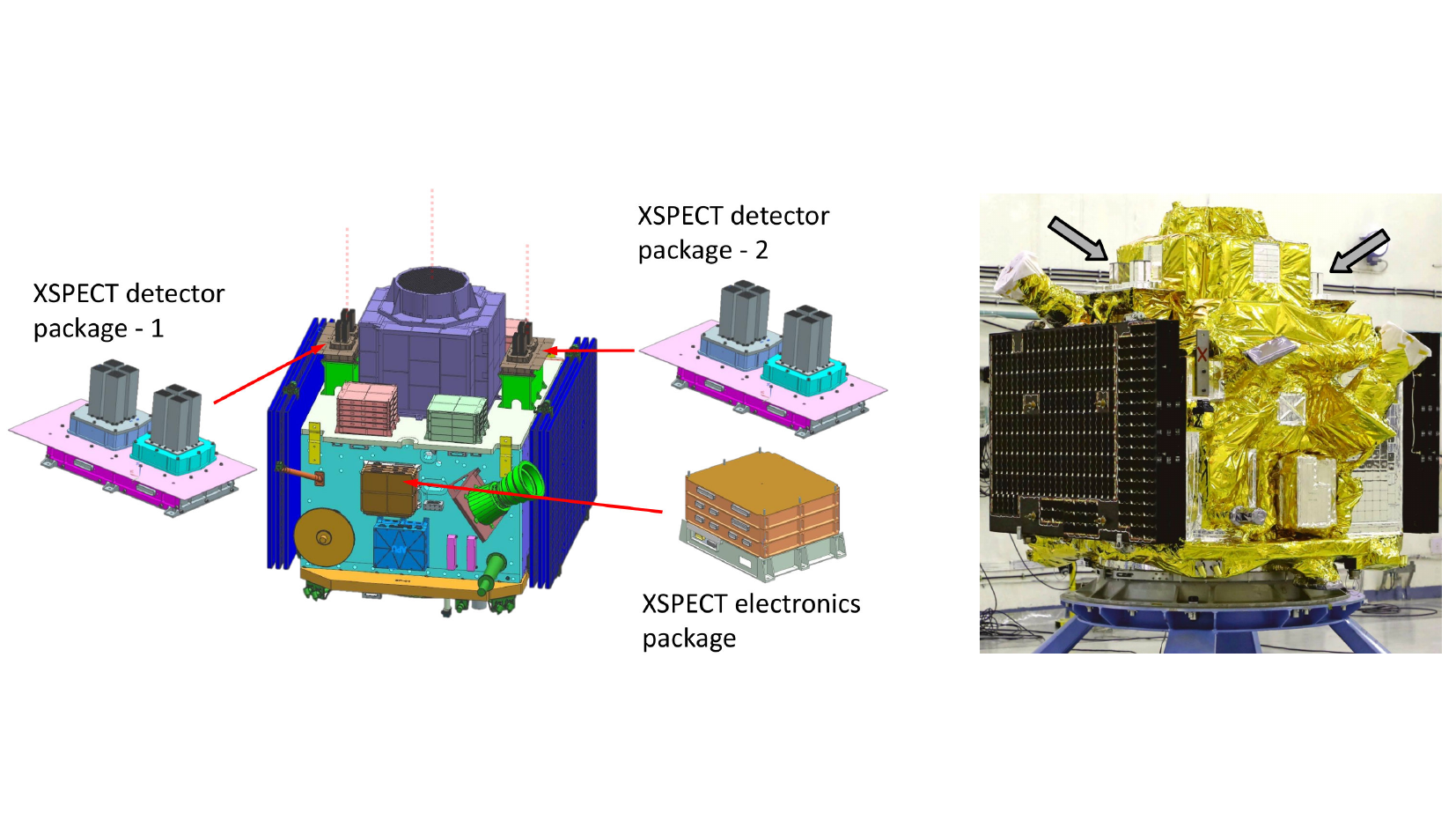}
\caption{(a) Layout showing the location of the components of XSPECT payload on-board XPoSAT. (b) Image of XPoSat after integration of all subsystems. The locations of the two \xsp detector packages are marked with arrows.}
\label{fig:xposat}
\end{figure*}

\xsp consists of sixteen~SCDs (numbered 0 to 15), grouped into four `quad' modules of four detectors each. A group of two quad modules consist of a detector package - two such packages are placed on either side of the POLIX instrument, on the spacecraft bus (Figure~\ref{fig:xposat}). Collimators are placed on top of the detectors to define the FOV of the detectors. The two quad modules of each detector package have collimators of two different square FOVs, viz. $2^{\circ}\times 2^{\circ}$ and $3^{\circ}\times 3^{\circ}$. Note that we use the terms `$2^{\circ}\times 2^{\circ}$' and `$3^{\circ}\times 3^{\circ}$' loosely to denote the two kinds of FOVs. The actual FOVs are slightly different, and are computed in Section~\ref{sec:arf}. The collimators are co-aligned to each other as well as to the POLIX axis, which defines the source pointing axis of the spacecraft. The choice of two different FOVs was made during instrument design for a better estimation of the local sky background. 

Since Si-based X-ray detectors are sensitive to optical light, each detector has an aluminised Polyimide optical light blocking filter placed below the respective collimators. Out of the sixteen detectors, one detector is covered by a Tantalum sheet of $500~\mu$m, to block soft X-rays and to obtain the local particle background, which mainly consists of Galactic Cosmic Rays (GCRs). Under ideal conditions, the data from two different FOVs, along with that from the blocked detector, can give a direct estimate of the background. More details can be found in Section~2.2 of [\citenum{rkrish2025}]. Table~\ref{tab:dettype} summarizes the different detectors along with their FOVs.

\begin{table}[h!]
\centering
\caption{Detector details}
\label{tab:dettype}
\begin{tabular}{cccc}
\hline
\textbf{SCD no.} & \textbf{Quad no.} & \textbf{FOV}                            \\ \hline
$0-3$                & 0 &  $\sim 2^{\circ}\times 2^{\circ}$ \\
$4-7$                & 1 &  $\sim 3^{\circ}\times 3^{\circ}$ \\
$8-11 $               & 2 &  $\sim 2^{\circ}\times 2^{\circ}$\\
$12-14$                & 3 &  $\sim 3^{\circ}\times 3^{\circ}$ \\
15                & 3 &  $\sim 3^{\circ}\times 3^{\circ}$ (blocked) \\\hline
\end{tabular}
\end{table}

\xsp operates in photon counting mode where an X-ray/particle interaction event may produce more than one sample above a pre-defined threshold. The details (viz. time, energy, detector ID, and split flag) of each sample above the threshold are recorded into data `packets'. The time stamping has a resolution of 1~ms. The split flag, an indicator of split events (see Section~\ref{sec:sinevt}), has a default value of 0, and is set to 1 whenever the previous and the present sample are above threshold. The packet duration is nominally 512~ms during source observations. However, while observing bright sources such as Sco~X-1, \xsp is operated in the lowest integration time mode of 256~ms packets, to maximize its count-rate handling capacity. The analog processing, digitisation, power conditioning, and data transfer with spacecraft interfaces is handled by the electronics package of \xsp. Further details on the \xsp instrument and its configuration can be found in our separate communication [\citenum{rkrish2025}]. 

The raw payload and spacecraft data (Level~0) downloaded at the ground stations is received at the Payload Operation Center (POC) at Space Astronomy Group, URSC where it is processed by automated pipelines to generate higher level (Level~1 and Level~2) scientific products [\citenum{rkrish2025}], and sent to the Indian Space Science Data Centre (ISSDC) for archival and dissemination. 

\section{On-ground Calibration}
\label{sec:ongroundcal}
To ensure the satisfactory performance of the instrument, a comprehensive pre-launch test and calibration activity was carried out, in addition to detailed simulations. This includes characterisation of the detector performance at different temperatures, charaterisation of the instrument spectral and timing response, verification of the various instrument parameters such as spetral resolution, field of view and alignment, estimate of the instrument noise and background levels etc. Basic ground tests and test results are explained in [\citenum{rkrish2025}]. In the following sections, some of those results are presented in more detail, in addition to other ground calibration aspects such as spectral response and noise characterisation. The validation and refinement of instrument calibration using on-board observations is discussed in Section~\ref{sec:onboardcal}.

\subsection{Performance verification}
\label{subsec:tvac}

\begin{figure*}
\centering
\includegraphics[scale=0.6, trim=1cm 2cm 1cm 2cm, clip=true]{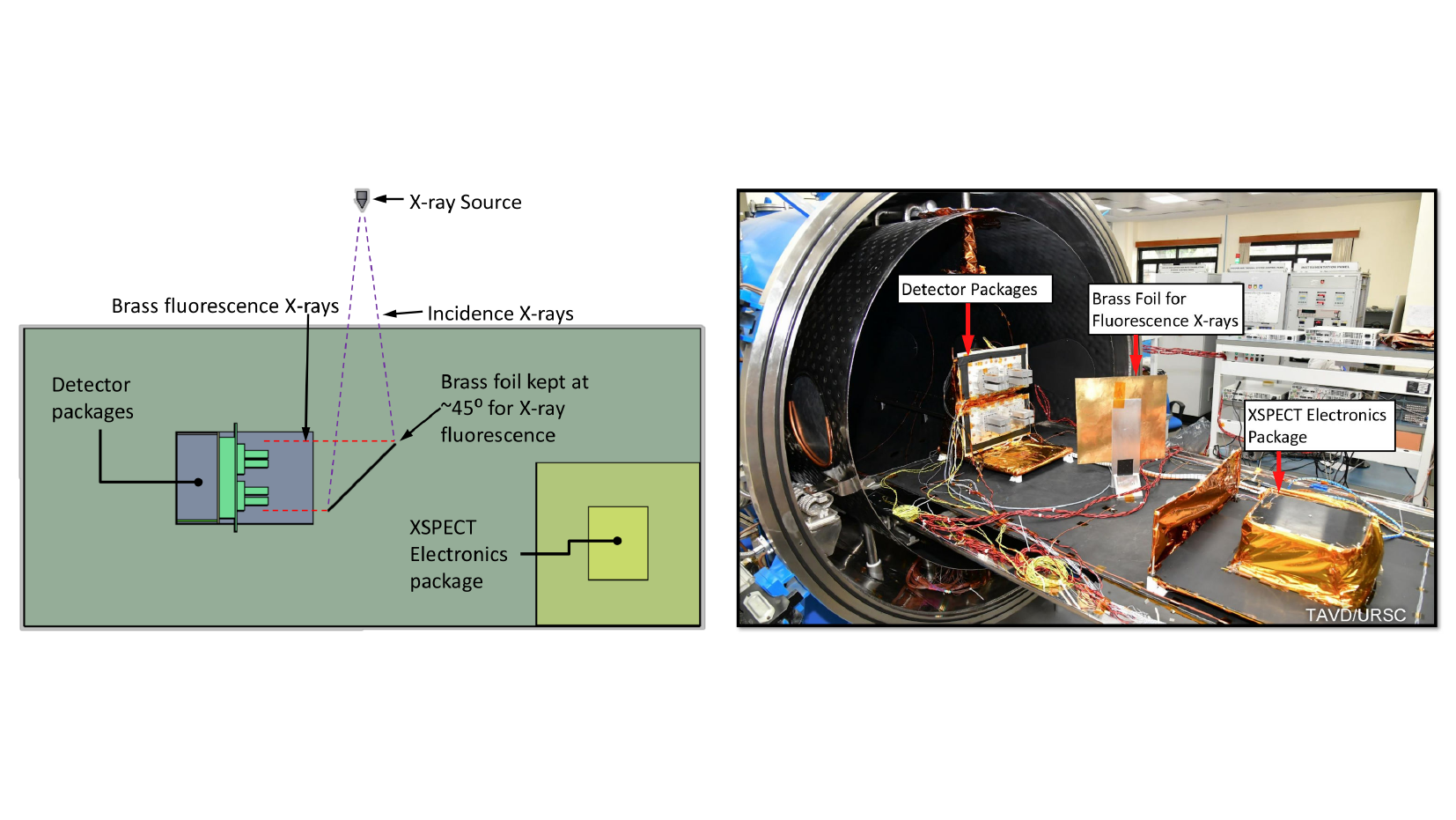}
\caption{(a) Schematic (top view) of the \xsp test setup inside the thermovacuum chamber. (b) Image of the placement of various components in the test chamber.}
\label{fig:tvacsetup}
\end{figure*} 

The fully assembled detector and electronics packages have undergone thermovacuum cycling on-ground, to qualify the subsystems for on-orbit conditions. The detector performance (gain and resolution) was monitored using fluorescence X-rays generated from a brass foil, kept at $\sim 45^{\circ}$ angle in front of the detector packages (Figure~\ref{fig:tvacsetup}). 

\par The thermovacuum cycling was carried out over the period of one week, subjecting the individual packages to several hot and cold cycles. In addition, the packages were also tested after complete integration on the spacecraft. System performance was found to be repeatable throughout the cycles, with the energy resolution well within the required specifications (FWHM $< 200$~eV @ 5.898 keV @ $-20^{\circ}$~C).

\subsection{Gain and FWHM}
\label{subsec:gainfwhm}
As mentioned earlier, SCDs operate in photon counting mode where each photon is individually detected as an `event', and each event produces an electrical signal proportional to the energy of the incoming photon. This signal (pulse height) is digitized by an ADC, whose value, for a given energy, is determined by the gain. The gain is dependent upon the detector temperature, and can have minor variations from detector to detector. Each quad has a thermistor, and the temperature reading is included in every data packet. 

\begin{figure*}
\centering
\includegraphics[scale=0.6, trim=0cm 1cm 0cm 1cm, clip=true]{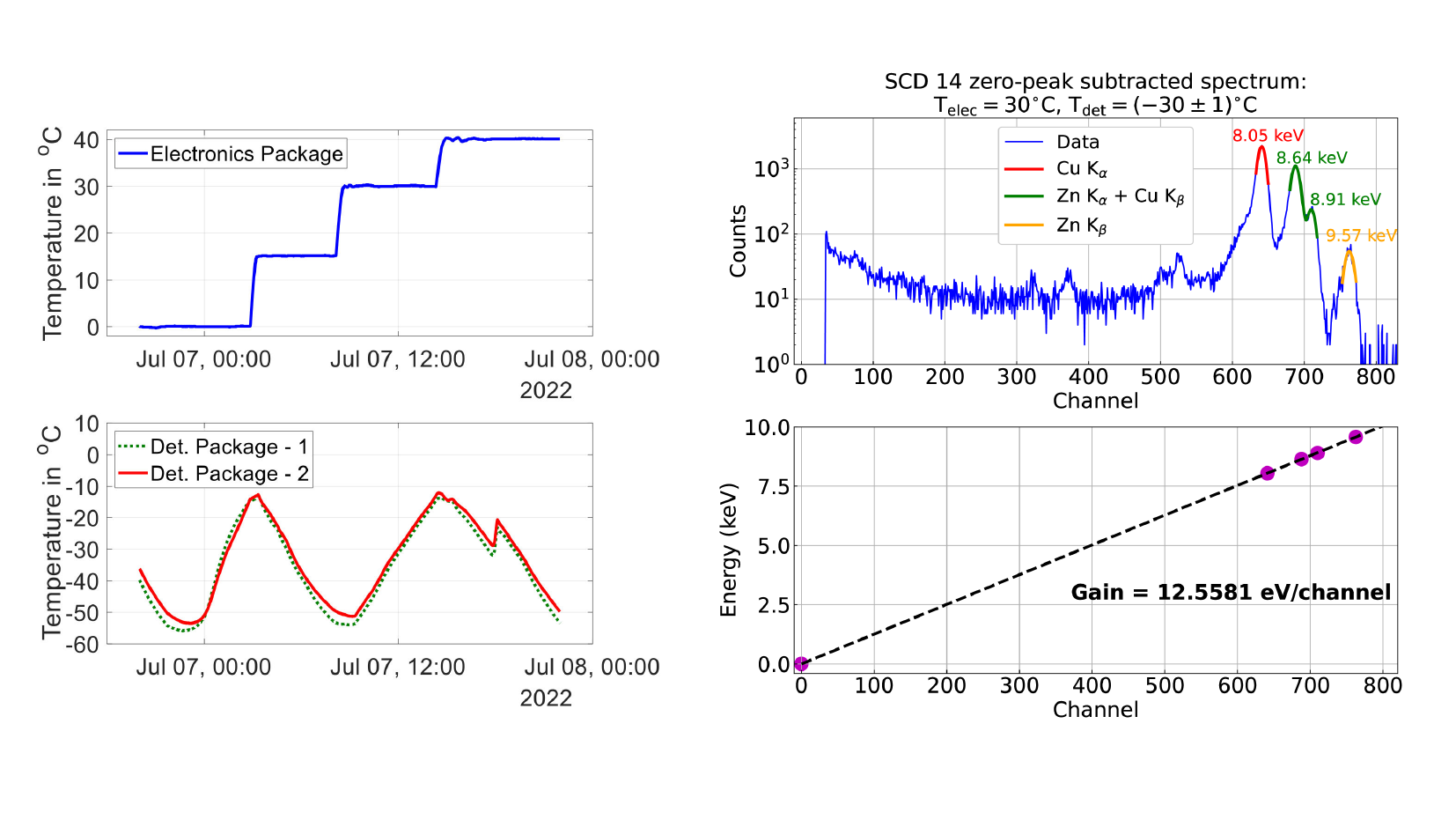}
\caption{(a) Temperature sweep of the (top) electronics and (bottom) detector packages. (b) (Top) Sample brass spectrum recorded by SCD 14 at the specified T$_{\mathrm{det}}$ and T$_{\mathrm{elec}}$, with the fluorescence peaks marked. (Bottom) Determined peak channels are fitted with a linear function to determine the gain. }
\label{fig:howgain}
\end{figure*}

The gain (in units of eV/channel) for each detector is determined experimentally by finding the channel of peak emission, when illuminated with standard X-ray energies. This is done using the same setup as shown in Figure~\ref{fig:tvacsetup}, by varying the system temperatures in a controlled manner. Both the detector and electronics packages are swept across the likely range of on-orbit temperatures for these subsystems (Figure~\ref{fig:howgain}a).

The spectrum is generated by first subtracting the pedestal level (`zero-peak' is the peak of the binned pedestal level, when there are no X-ray interactions) from the digitized channel value of each event, and then binning them. Subsequently, the various peaks in the brass fluorescence spectrum are fitted with Gaussians, and the peak channels along with their known energies are then fitted with a linear function to determine the gain of the device, as illustrated in Figure~\ref{fig:howgain}b.

\begin{figure*}
\centering
\includegraphics[scale=0.6, trim=0cm 1cm 0cm 1cm, clip=true]{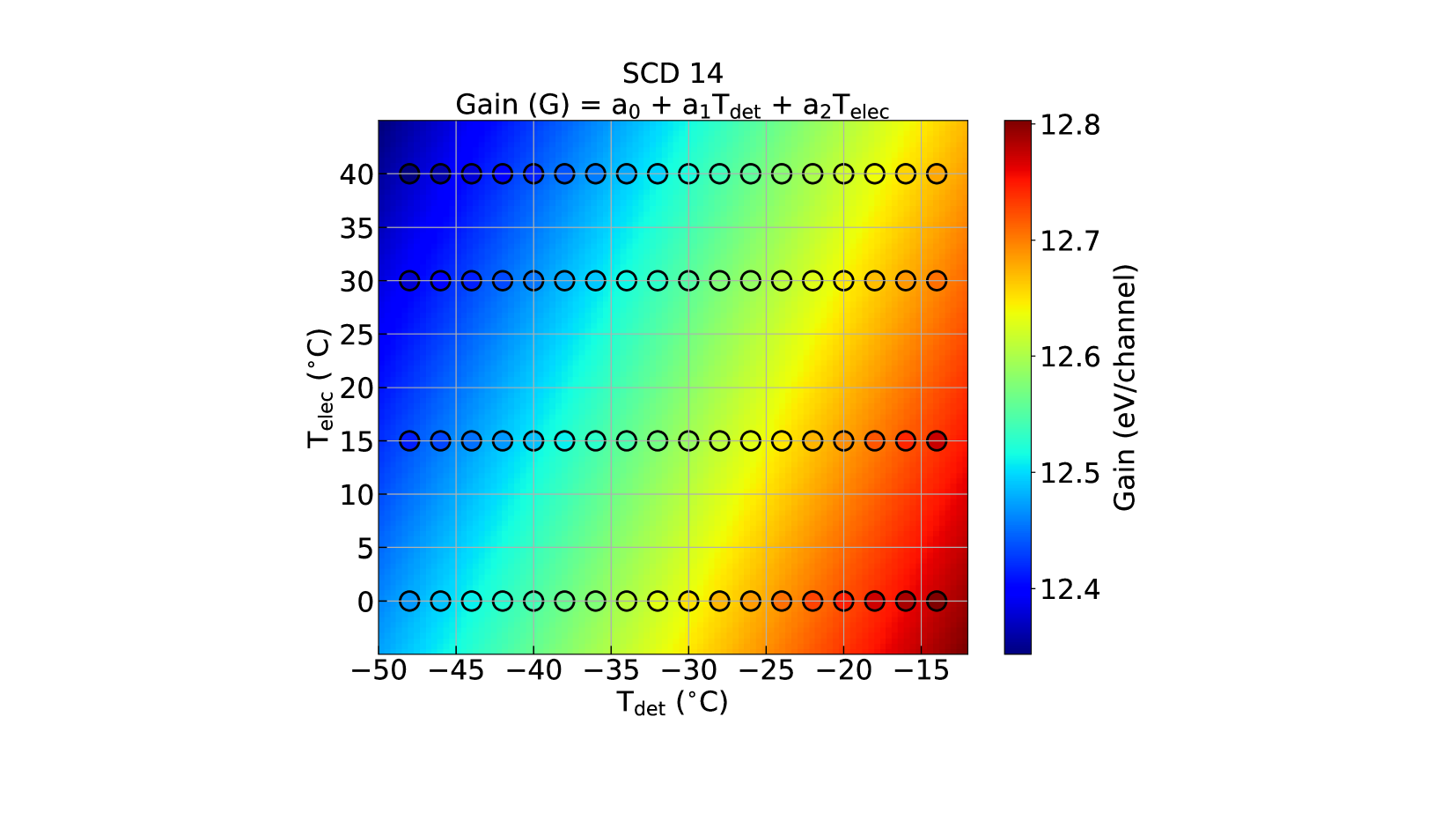}
\caption{Two-dimensional fit to the variation of gain as a function of T$_{\mathrm{det}}$ and T$_{\mathrm{elec}}$ for SCD~14. The markers represent the data points with their colors representing the measured values of gain. The background colormap represents the surface generated using the best-fit parameters of equation (1) for SCD~14.}
\label{fig:gain2d}
\end{figure*} 

The gain is found to depend linearly on both detector temperature (T$_{\mathrm{det}}$) and electronics temperature (T$_{\mathrm{elec}}$), with a positive relation to the former and a negative relation to the latter. Hence the gain variation for each detector is fitted with a 2-dimensional function of the form
\begin{equation}
G_i (T_{\mathrm{det}}, T_{\mathrm{elec}}) = a_{0i} + a_{1i}T_{\mathrm{det}} + a_{2i}T_{\mathrm{elec}}
\end{equation} 
where the coefficients $a_{0i}$, $a_{1i}$ and $a_{2i}$ for detector $i$ are obtained from the fit. This is shown in Figure~\ref{fig:gain2d} for SCD~14 using a colormap. The set of coefficients is part of the calibration database which is used to assign energy to each event based on the temperatures. The average gain offset ($a_0$) across the fifteen detectors is found to be $\sim 12.5$~eV/channel, which is used to define the pulse invariant (PI) channels for \xsp spectra. PI channel number represents the detected photon's energy that has been corrected for instrument gain, ensuring energy assignments remain stable even under changing detector conditions. The average values of $a_1$ and $a_2$ are, respectively, 0.0084~eV~ch$^{-1}$~$^{\circ}C^{-1}$ and $-0.0025$~eV~ch$^{-1}$~$^{\circ}C^{-1}$.

\begin{figure*}
\centering
\includegraphics[scale=0.5]{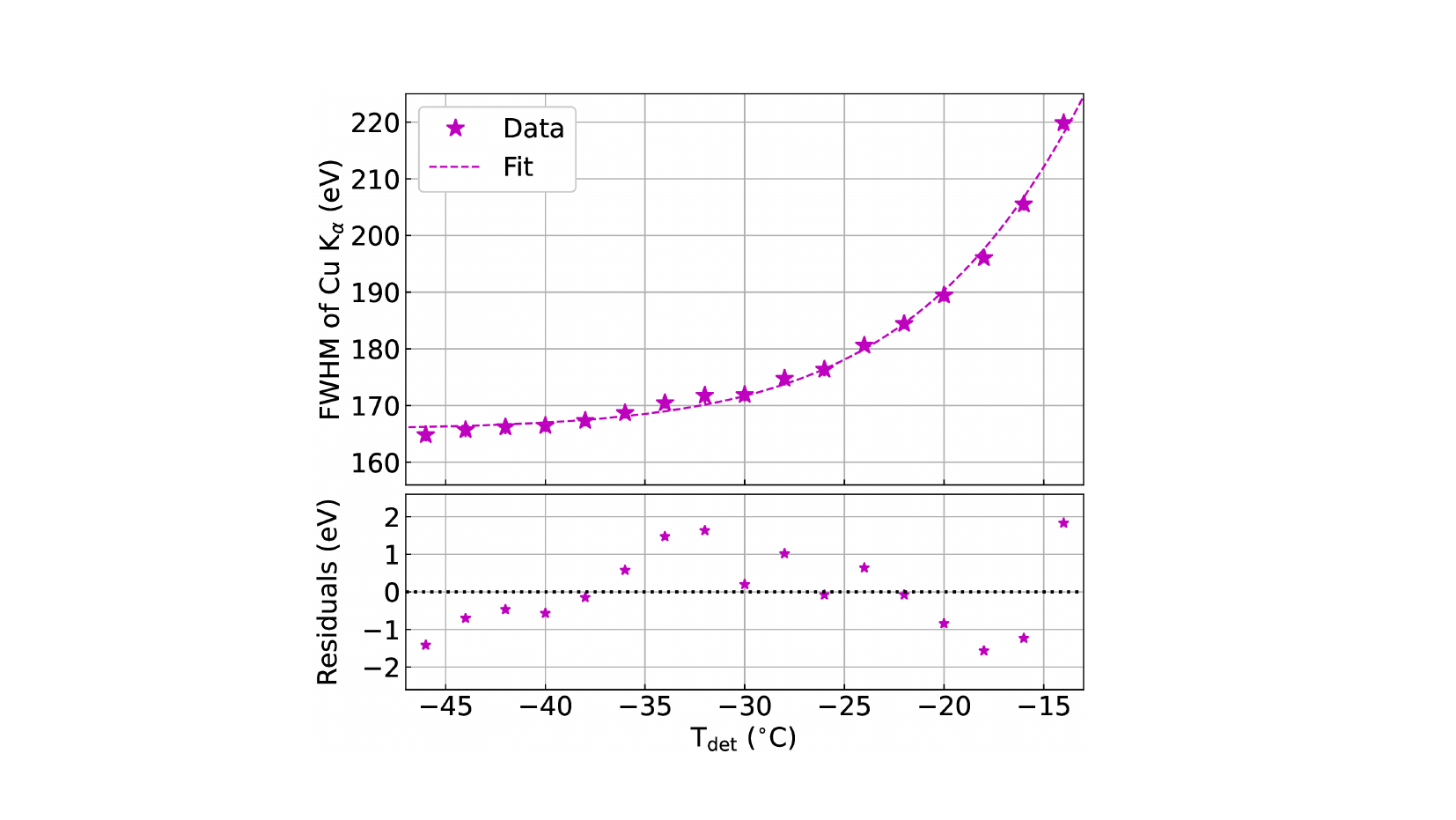}
\caption{(Top) Variation of FWHM with T$_{\mathrm{det}}$. Spectra have been added across all the detectors. The dashed line is the best fit to the data using Eq. (2). (Bottom) Residuals of the best fit.}
\label{fig:fwhmvst}
\end{figure*} 
The variation of the energy resolution (FWHM of Cu K$_\alpha$ line) as a function of T$_{\mathrm{elec}}$ is found to be negligible and thus we only consider the variation with respect to T$_{\mathrm{det}}$. The spectra from all detectors within a specific T$_{\mathrm{det}}$ bin ($\Delta T_{\mathrm{det}}=2^{\circ}C$) are combined, and the resultant FWHM variation of the Cu K$_{\alpha}$ peak is shown in Figure~\ref{fig:fwhmvst}. The temperature-dependent leakage current is proportional to $T^{3/2}\exp\left(\frac{-E_g}{2kT}\right)$, where $E_g$ is the Silicon band gap energy, $k$ is the Boltzmann constant, and $T$ is the detector temperature in Kelvin. Combining this with the temperature-independent terms, we obtain
\begin{equation}
FWHM = \sqrt{a+bT^{3/2}\exp\left(\frac{-c}{T}\right)}
\end{equation}
where $a$, $b$ and $c$ are constants (see [\citenum{janesick2001}]). $a$ represents the temperature-independent component of the $FWHM$, which includes on-chip noise and statistical (fano) noise [\citenum{knoll2000}]. $b$ is a normalization for the temperature-dependent term, and $c$ represents the constant factor $E_g/2k$. The FWHM variation of our devices is well-fitted with this equation, giving $a = 26002.8$, $b = 1.0\times 10^{10}$, and $c = 5573.6$. At very low temperatures, the dark current term is neglible, and hence $\sqrt{a}$, which is the asymptotic value, represents the noise performance of the device.

\subsection{System noise and Fano factor}
\label{subsec:fano}

\begin{figure*}
\centering
\includegraphics[scale=0.5, trim=0cm 3.5cm 0cm 2cm, clip=true]{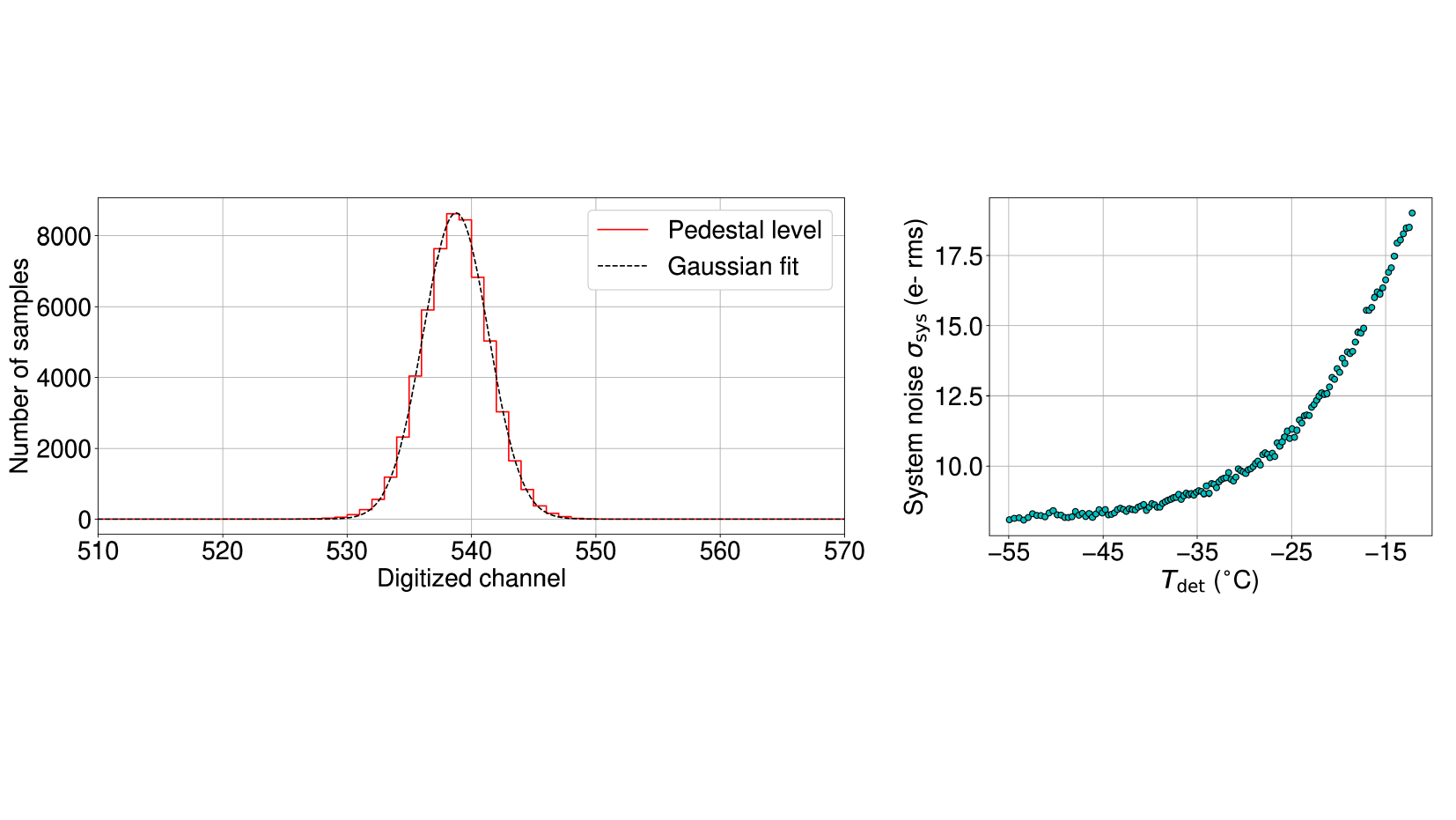}
\caption{Noise characterization of SCD~14: (a) fitting of the histogram of pedestal level samples ($T_{\mathrm{det}}\sim -41^{\circ}$C) to obtain the noise, and (b) variation of system noise ($\sigma_{\mathrm{sys}}$) in electrons (rms) as a function of $T_{\mathrm{det}}$}
\label{fig:noise}
\end{figure*}

One of the modes of \xsp writes the pedestal events to the data packet, hence allowing an estimation of the noise of the system. To characterise the noise performance, we obtained data during ground tests in this mode across a range of T$_{\mathrm{det}}$ which is expected to affect the dark current. The data is then segregated by T$_{\mathrm{det}}$, and the obtained zero-peak is fitted by a Gaussian function to characterise the system noise ($\sigma_{\mathrm{sys}}$) as a function of detector temperature. This is shown in Figure~\ref{fig:noise} for one of the devices. $\sigma_{\mathrm{sys}}$ consists of the on-chip electronics noise and the temperature-dependent dark current. For the range of temperature swings observed on-board ($\sim -49^{\circ}C$ to $\sim -32^{\circ}C$), the average $\sigma_{\mathrm{sys}}$ across all the devices is found to be $\sim 10~e^-$~rms.

\par The Fano factor ($F$), which depends on the type of semiconductor, is a measure of the actual variance in the number of electron-hole pairs produced in the detector by ionizing radiation (quantifying its departure from a pure Poisson distribution), and determines the best possible energy resolution achievable by a semiconductor device. Many authors have experimentally determined the Fano factor for Si [e.g. \citenum{rodrigues2021, kotov2018, lowe2007}], as well as predicted it theoretically [\citenum{subashiev2010, alig1980}] or using simulations [\citenum{fraser1994, mazziotta2008, brigida2004}], and several values are found in literature, ranging from $\sim 0.08$ [\citenum{jordan2008, eberhardt1970}] to $\sim 0.16$ [\citenum{perotti1999}]. $F$ can be calculated using the relation [\citenum{janesick1988, owens2002}]
\begin{equation}
F = \frac{\left(\frac{FWHM_{T,E}}{2.35}\right)^2 - \sigma_{\mathrm{sys, T}}^2}{wE}
\end{equation}
where $w$ is the mean ionisation energy (3.65~eV for Si). Using the Cu~K$_{\alpha}$ ($E=8.05$~keV) FWHM and the computed values of $\sigma_{\mathrm{sys, T}}$ as explained above, the average value of $F$ is observed to be 0.149. 

\subsection{Spectral redistribution function}
\label{subsec:rrcat}
For inversion (deconvolution) of spectral data, accurate knowledge of the instrument spectral response is essential. To obtain the spectral redistribution function (SRF) of \xsp, we carried out experiments at Raja Ramanna Center for Advanced Technologies (RRCAT), Indore. This is a synchrotron facility where monochromatic X-ray beams are available at different beamlines. We used two beamlines to cover the \xsp energy range: BL-03 for the lower energies (0.5 to 1.6 keV) and BL-16 for the higher energies (6 to 16 keV). We acquired spectra at every 100~eV between $0.5-1.6$~keV and at every 1~keV between $6-16$~keV, with an additional acquisition at 6.5~keV. For the mid-range energies between 1.6 and 6 keV, fluorescence experiments with Ta, Au, Cl, Ca, and Ti using monochromatic excitation from the X-ray beam were carried out. The detailed experimental setup as well as method of SRF generation will be covered in a separate paper. Here, we only mention the results.

\begin{figure*}
\centering
\includegraphics[scale=0.6, trim=2cm 2cm 1cm 2cm, clip=true]{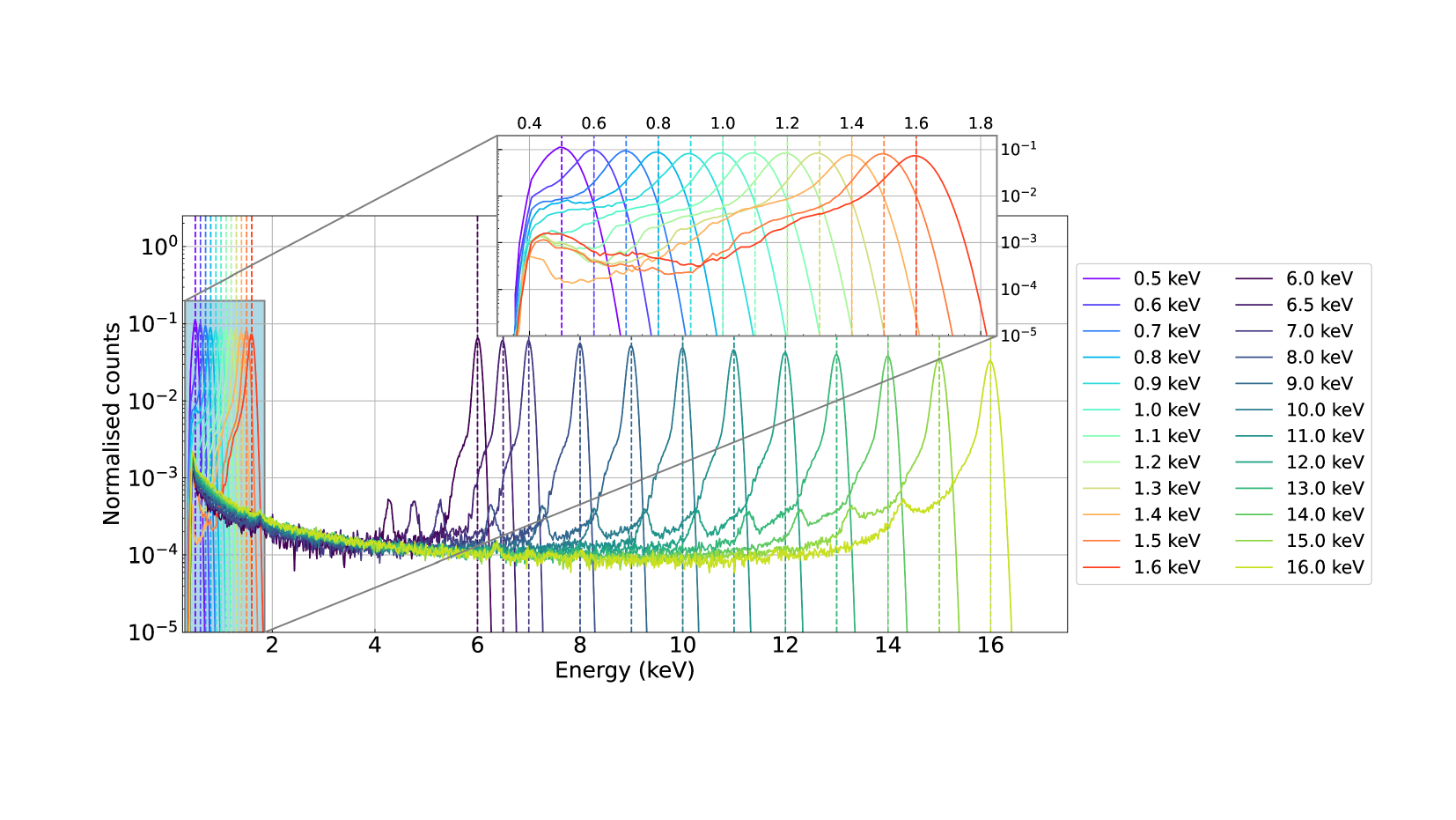}
\caption{Normalised spectra of monochromatic energies obtained at RRCAT. The inset shows the zoomed-in view of the lower energy spectra.}
\label{fig:srf_mono}
\end{figure*} 

\par Since the SRF is not expected to vary from device to device of the same batch, we carried out the experiments on a single SCD, belonging to the same batch of devices as those used in the flight model of \xsp. The electronics threshold for event selection (above the noise level) was set to $\sim 0.5$~keV, which is same as the threshold to be used on-board. During the experiments, the SCD was maintained at a constant temperature of $\sim -35^{\circ}$C, and a calibration dataset using Fe-55 radioactive source was taken before each experiment. We applied event filtering logic to discard the split events and generated single event spectra from our data (Section~\ref{sec:sinevt}). Figure~\ref{fig:srf_mono} shows the (normalised) spectra obtained from the monochromatic experiments.  

\begin{figure*}
\centering
\includegraphics[scale=0.5, trim=4cm 2cm 4cm 2cm, clip=true]{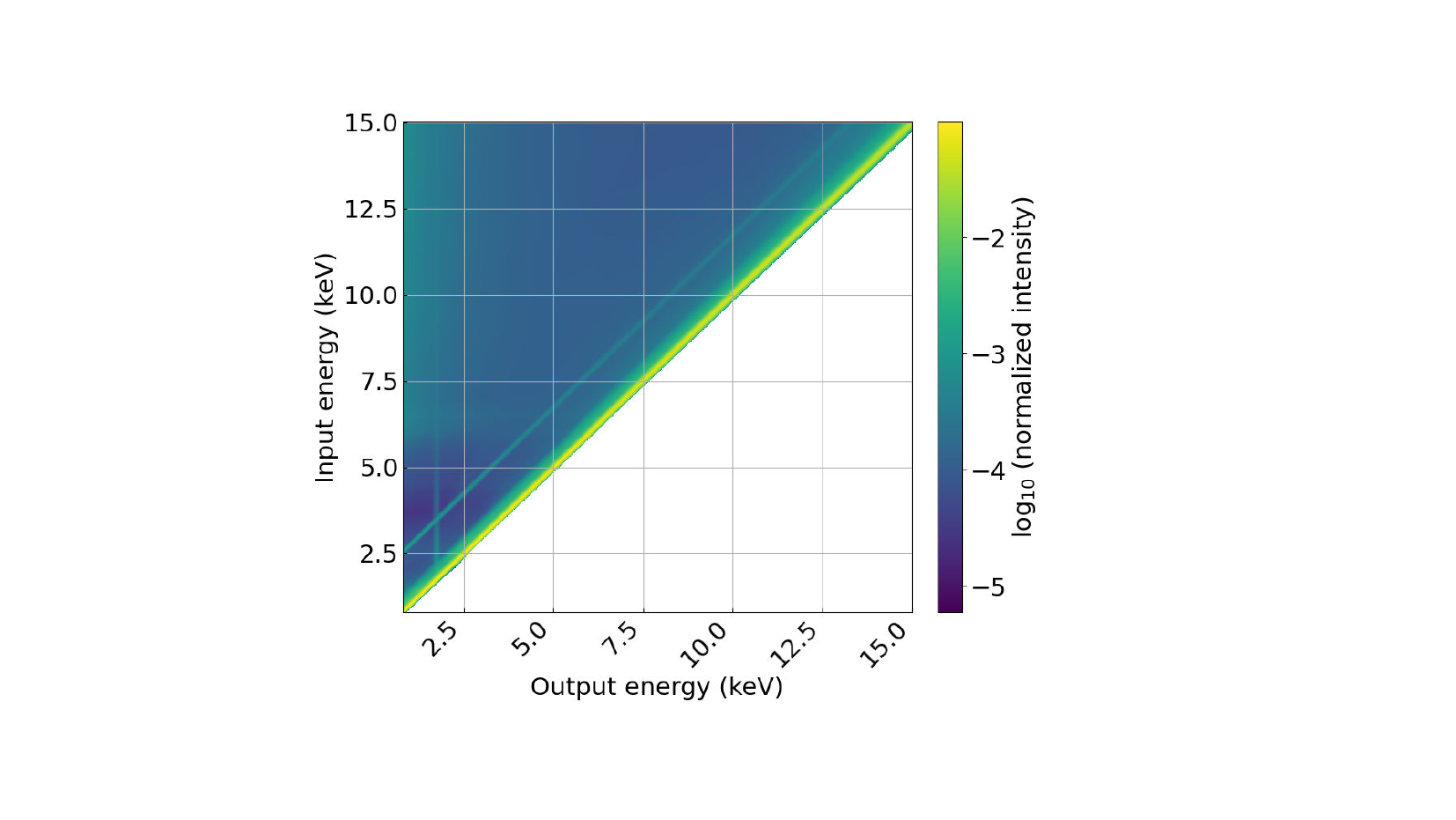}
\caption{\xsp $0.8-15$~keV response matrix, in logarithmic units.}
\label{fig:fullsrf}
\end{figure*} 

The final response is generated by a weighted interpolation method, with the weights inversely proportional to the difference ($|\Delta E|$) of the target energy (to be interpolated) and the parent energy (observed). This method has been used to generate the response matrix in $0.5-16$~keV (in steps of 10~eV). Figure~\ref{fig:fullsrf} shows the interpolated response matrix, in which the bright diagonal trace represents the photopeak, and the fainter line above it represents the Si escape peak. The faint vertical line $\sim 1.7$~keV is the Si fluorescence from the detector. 

\subsection{Single event correction factor}
\label{sec:sinevt}
\begin{figure*}
\centering
\includegraphics[scale=0.6, trim=1cm 1cm 0cm 1cm, clip=true]{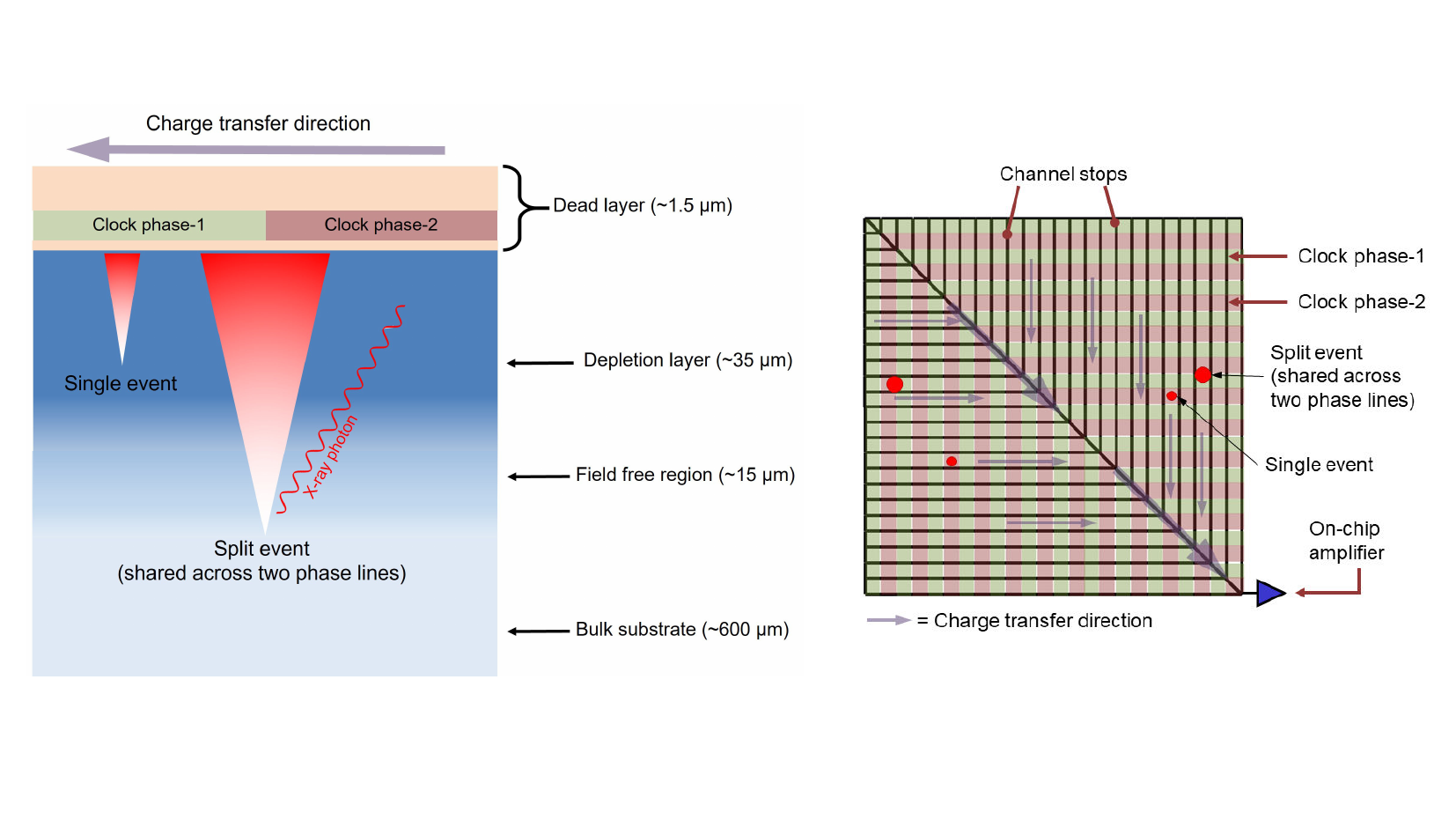}
\caption{(a) Schematic (not to scale) of the cross-section of SCD. The approximate layer thicknesses are taken from [\citenum{smith2014}]. (b) Schematic of the top view of SCD, showing its readout scheme. Events contained within a single pixel (single event) are fully read out in a single clock cycle, whereas those shared across multiple pixels (split events) are read out in two or more successive cycles.}
\label{fig:scd_structure}
\end{figure*}

The vertical structure of SCDs, which are front-illuminated devices, consists of various layers, viz. top dead layers, depletion region (region of complete charge collection), field-free region (region of partial charge collection), followed by a bulk substrate, as shown in Figure~\ref{fig:scd_structure}a. During vertical charge transport from various depths of the detector to the buried channel, it experiences diffusion which may lead to the charge cloud being shared across multiple `pixels', as depicted in Figure~\ref{fig:scd_structure}a. In addition, charge sharing can also happen for events occurring at the boundary of two pixels. During the `sweeping' readout process [\citenum{lowe2001}], the deposited charge is clocked towards the diagonal node, as shown by the arrows in Figure~\ref{fig:scd_structure}b. This produces a sequence of samples at readout speed (100~kHz). A `single' (or isolated) event is defined when n$^{\mathrm{th}}$ sample is above the energy threshold (defined as 0.5~keV, [\citenum{rkrish2025}]) with (n-1)$^{\mathrm{th}}$ and (n+1)$^{\mathrm{th}}$ samples below the energy threshold. When two or more consecutive samples are above the threshold, we define it as a `split' event. This is similar to grading of events in a CCD X-ray imager [e.g. \citenum{ascaabc,chandrapog,xmm}].

In a `raw' spectrum (all events considered, irrespective of splitting), the contribution of continuum to total counts is undesirably higher than in the case of a spectrum constructed using only single events ($\sim$38\% for all vs $\sim$12\% for single, as estimated from monochromatic spectra at 6~keV). Additionally, energetic particle interactions also give rise to split events. For example, the ratio of all events to single events in the spectrum of the blocked detector (which is predominantly due to particles) is $\sim 5$. Selecting single events for spectral analysis thus serves a dual purpose: reduction of continuum in the SRF and reduction of particle background. Hence, for scientific analysis, we recommend using only the single events. All scientific data products (default spectra and lightcurves) as well as responses and background files are filtered to include only single events.

The fraction of single events varies as a function of energy, and this was computed using the monochromatic data obtained at RRCAT (Section~\ref{subsec:rrcat}). Subsequently, the trend was interpolated to be included in the device quantum efficiency, and is shown in Figure~\ref{fig:arftune}b.

\section{On-board Calibration and First Results}
\label{sec:onboardcal}

In this Section, we discuss the initial on-board observations by \xsp, and describe how they were used to verify and refine the ground calibration. These observations were primarily carried out during the PV phase as mentioned in Section~\ref{sec:intro}. Post initial system checks, XPoSat commenced source observations on 5$^{\mathrm{th}}$~January~2024 with the supernova remnant (SNR) Cassiopeia-A being the first target. Throughout the PV phase, various standard sources such as SNRs Cas-A and Tycho, Crab pulsar, GX 301-2, etc were observed. In addition to source pointing observations, offset observations as well as scan observations of the Crab pulsar were carried out. To generate the background spectrum, several blank sky regions were also observed.

\subsection{Data processing}
For obtaining the results presented in the following sections, the Level~0 data from the instrument is first processed to generate spectra and light curves (Level~2 products). Since the method of data processing not only impacts the results directly but also determines the quality of calibration, the main steps involved in the Level~0 - Level~1 - Level~2 processing are described here in brief. A description of the different levels of data can be found in [\citenum{rkrish2025}].
\begin{itemize}
\item L0 - L1 conversion:
	\begin{itemize}
		\item Raw binary data recorded by the payload is segregated into event data and instrument housekeeping (HK) data.
		\item Event time stamps in the data packet are in instrument frame, which are transformed to UTC by correlating with data from an on-board GPS clock receiver.
		\item The temperature-dependent gains are applied to each event and subsequently assigned pulse-invariant (PI) channels. The detector and electronics box temperatures are available as part of HK data.
		\item The attitude and orbit files are generated (at 1~s cadence, nominally) from spacecraft SPICE kernels [\citenum{spice}].
		\item A filter file is generated which contains all the parameters necessary for applying any kind of filtering to the events. The columns in the filter file are generated using the relevant columns from the attitude, orbit, instrument HK and auxiliary HK (containing spacecraft-related parameters) data.
		\item Nominal screening criteria (Table~\ref{tab:stdgti}) are applied to generate the standard good time interval (GTI) file. Note that many of the default values are as per on-ground expectations of performance, and may be modified in the future with evolving understanding and performance of the instrument.  
	\end{itemize}
	
	\begin{table}[h!]
	\caption{Filtering criteria used to generate the standard GTI}
	\label{tab:stdgti}
	\resizebox{\columnwidth}{!}{%
	\begin{tabular}{ccc}
	\hline
	\textbf{Paramter}                & \textbf{Default value/range }                    & \textbf{Remarks}                                            \\ \hline
	Elevation               & $> 5^{\circ}$ & Angle of S/C pointing direction above bright earth          \\
	Sun angle               & $ > 45^{\circ}$                         & Angle between Sun and S/C pointing direction                \\
	Moon angle              & $> 10^{\circ}$                          & Angle between Moon and S/C pointing direction               \\
	SAA flag                & $=0$                                       & 0: not in SAA, 1: in SAA                                    \\
	Time since SAA          & $> 100$ seconds                         & Time elapsed after exitting SAA                             \\
	Detector temperature    & $\epsilon \;[-70^{\circ}C, -15^{\circ}C]$ & As per instrument operation range \\
	Electronics temperature & $\epsilon \;[0^{\circ}C, 40^{\circ}C]$     & As per instrument operation range \\
	Angular offset          & $<0.1^{\circ}$                          & Angle between source coordinates and S/C pointing direction \\
	Is night flag           & $=1$                                      & 0: day , 1: night (crossed the terminator to eclipse side)  \\
	Earth visible flag           & $=0$                                      & Whether earth is visible in the FOV (0: no, 1: yes)  \\
	
	 \hline
	\end{tabular}
	}	
	\end{table}

\item L1 - L2 conversion:
	\begin{itemize}
		\item Standard GTI is applied to the events to produce a screened event file.
		\item The FOV-wise (see Table~\ref{tab:dettype}), $0.8-15$~keV single event light curves (10~s binsize) and spectra are generated from the screened events. 
	\end{itemize}
	
\end{itemize}

A dedicated XSPECT data processing software package, which includes routines for the regeneration of L2 products from L1 data using custom criteria, as well as some useful utilities, has been developed in Python, and is made available for the user community.

\subsection{Verification of gain calibration and spectral performance}
SNRs Cas-A and Tycho are standard calibration sources used by soft X-ray telescopes, due to the presence of multiple emission lines of ionized elements in the energy band of interest. Owing to the thermal design of \xsp, orbital temperature swings typically remain between $-49^{\circ}$C to $-32^{\circ}$C, which is well within the calibrated temperature range of the detectors. The detector-wise temperature-dependent gain calibration obtained on-ground is applied to the on-board data, hence the observed line energies from the SNRs can be used to confirm the correctness of the gain values.  Moreover, comparing the observed line widths with the expected values (derived from the on-ground FWHM numbers) will provide confidence on the calibration. 

\begin{figure*}
\centering
\includegraphics[scale=0.61, trim=0cm 0cm 0cm 0cm, clip=true]{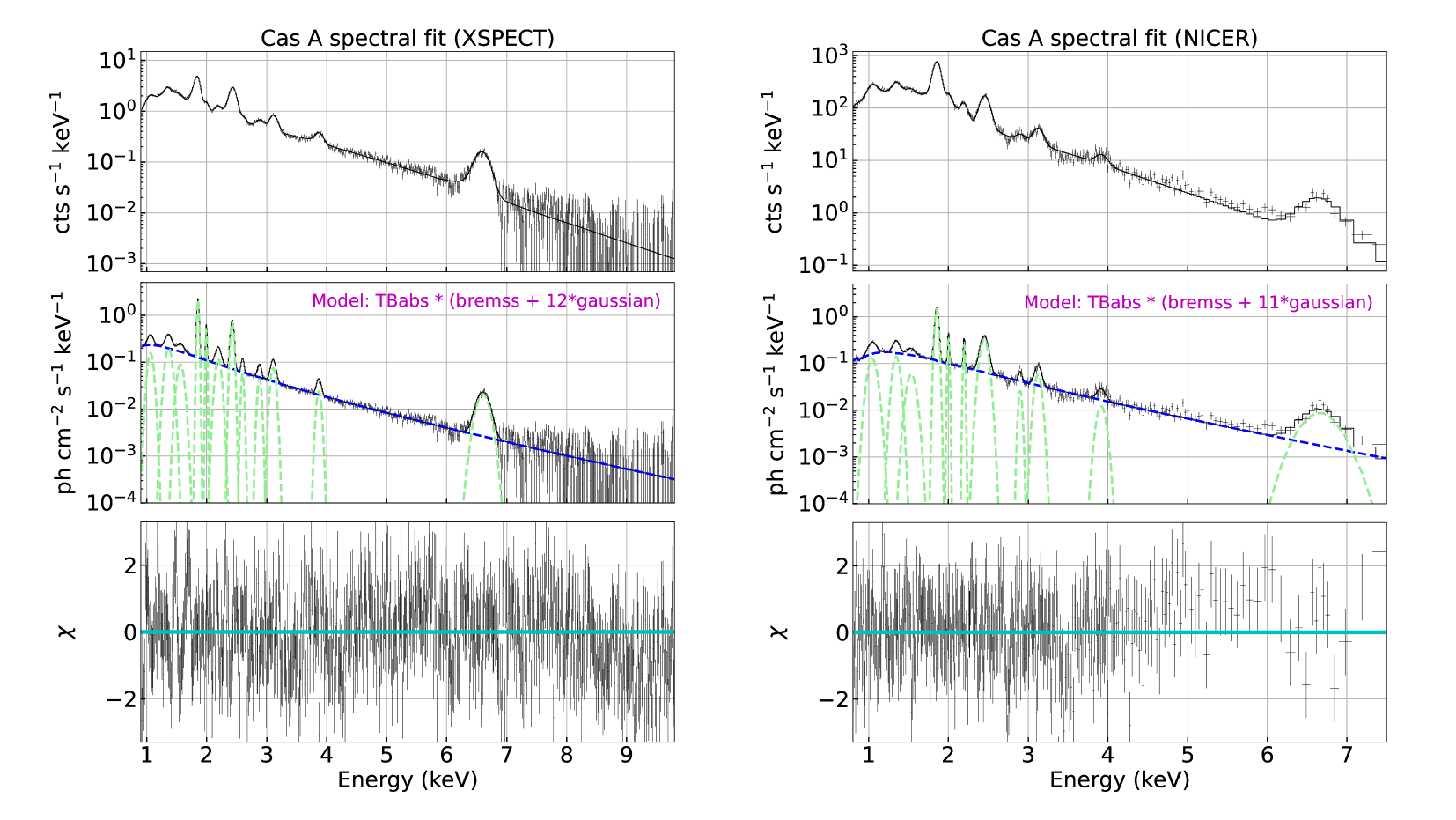}
\caption{Cas~A spectral fitting using (a) \xsp and (b) \textit{NICER} data. In both cases, the top, middle and bottom panels show the (background-subtracted) observed spectrum, unfolded spectrum and residuals (in units of $\sigma$) respectively.}
\label{fig:SNR_cas}
\end{figure*} 

\begin{table}
\centering
\caption{Best fit spectral parameters of supernova remnants. Data used is XSPECT and NICER for Cas A, and XSPECT and SXT for Tycho. The quoted flux is the unabsorbed flux in $0.5-10$~keV, in units of ergs~s$^{-1}$~cm$^{-2}$. See text for details}
\label{tab:snrfits}
\begin{tabular}{ccccc}
\hline
\multirow{2}{*}{\textbf{Parameter}} & \multicolumn{2}{c}{\textbf{Cas A}}  & \multicolumn{2}{c}{\textbf{Tycho}} \\ \cline{2-5} 
                                    & \textbf{XSPECT} & \textbf{NICER}    & \textbf{XSPECT}   & \textbf{SXT}   \\ \hline
nH ($\times 10^{22}$ cm$^2$)        & $0.316\pm0.021$   & $0.507\pm0.015$     &    0.1   &     0.1    \\
kT$_{\mathrm{bremss}}$ (keV)        & $2.060\pm0.032$   & $1.787\pm0.060$     &      $1.02\pm0.05$  &   $1.22\pm0.09$ \\
norm$_{\mathrm{bremss}}$            & $0.727\pm0.022$   & $0.813\pm0.047$     &  $0.208\pm0.008$  & $0.281\pm0.029$                  \\\hline
$E_1$ (keV)    & - &    -    &  $0.855\pm0.009$    &  $0.832\pm0.004$ \\
$E_2$ (keV)                         & $1.060\pm0.004$ & $1.037\pm0.003$       &  $1.082\pm0.013$    &  $1.099\pm0.012$ \\
$E_3$ (keV)                         & $1.365\pm0.004$ & $1.344\pm0.000$   &         $1.347\pm0.010$ &    $1.312\pm0.007$            \\
$E_4$ (keV)                         & $1.564\pm0.010$ & $1.533\pm0.014$   &       -               &   -             \\
$E_5$ (keV)                         & $1.857\pm0.001$ & $1.854\pm0.001$   &      $1.865\pm0.002$    &               $1.863\pm0.002$ \\
$E_6$ (keV)                         & $1.997\pm0.002$ & $2.002\pm0.003$   &      $1.959\pm0.017$    &               $1.796\pm0.011$ \\
$E_7$ (keV)                         & $2.196\pm0.003$ & $2.198\pm0.003$   &      $2.216\pm0.007$    &               $2.218\pm0.005$ \\
$E_8$ (keV)                         & $2.428\pm0.001$ & $2.445\pm0.002$   &       $2.442\pm0.002$   &               $2.462\pm0.002$ \\
$E_9$ (keV)                         & $2.601\pm0.006$ & -                 &            -            &               - \\
$E_{10}$ (keV)                         & $2.879\pm0.005$ & $2.898\pm0.012$   &      $2.887\pm0.014$    &               $2.909\pm0.010$ \\
$E_{11}$ (keV)                        & $3.105\pm0.002$ & $3.122\pm0.005$   &    $3.097\pm0.007$    &               $3.132\pm0.008$ \\
$E_{12}$ (keV)                        & $3.867\pm0.022$ & $3.916\pm0.014$   &    $3.862\pm0.021$    &               - \\
$E_{13}$ (keV)                        & $6.599\pm0.004$ & $6.656\pm0.021$   &          -        &                \\\hline
$\chi^2_{\mathrm{red}}$             & 1.42 (983/692) & 1.17 (402/344)    &    1.45 (395/272)     &               1.41 (415/295) \\ 
Flux ($\times 10^{-9}$)     & $2.67\pm0.04$ & $2.49\pm0.05$    &    $0.55\pm0.051$                        &       $0.60\pm0.04$         \\ \hline
\end{tabular}
\end{table}

The top panel of Figure~\ref{fig:SNR_cas}a shows the observed spectrum of Cas~A from the $2^{\circ}\times2^{\circ}$ detectors after background subtraction (Section~\ref{sec:bkg}), for an exposure time of $\sim 86$~ks. The spectrum in the range $0.8-10$~keV (spectrum is background-dominated beyond $\sim8$~keV) was fitted with a model consisting of an absorbed bremsstrahlung component, along with twelve Gaussian lines. For four of the lines (1.4~keV, 1.6~keV, 2.0~keV and 2.6~keV), the line width ($\sigma$) could not be constrained when left free. Hence, they were frozen (50~eV, 50~eV, 10~eV and 20~eV respectively) to values which minimized the residuals. Higher widths can be attributed to the presence of unresolved line complexes at these energies. For all other lines, the line width was constrained in the range $\sim15-95$~eV. We obtain a good fit using this model, with a $\chi^2_{\mathrm{red}}$ of 1.4. The best fit parameters are summarized in Table~\ref{tab:snrfits}. To validate the fitting, we also fitted the Cas~A spectrum observed by \textit{NICER} [\citenum{gendreau2016}] (observation~ID 6010080284) using the same model, which is shown in Figure~\ref{fig:SNR_cas}b. The closely matching line energies of \xsp and \textit{NICER} (Table~\ref{tab:snrfits} columns~2 and 3) as well as consistency with published results [e.g. \citenum{holt1994}] validate the gain coefficients determined from ground calibration (Section~\ref{subsec:gainfwhm}). Periodic calibration observations are planned throughout mission life to monitor these coefficients, and update them if required.

Similarly, the spectrum of SNR Tycho was fitted with an absorbed bremsstrahlung model plus ten Gaussian lines. For comparison, we also fitted the Astrosat/SXT [\citenum{singh2017}] spectrum using the same model. The line energies obtained from the spectral fits of both the instruments are tabulated in Table~\ref{tab:snrfits} (columns~4 and 5) and again, a good match is observed.

Since the intrinsic width of isolated lines from SNRs is small, the width of the observed lines can be used to set an upper limit on the on-board spectral performance. For example, the fitted FWHM of the identified Si line in Cas~A at $1.86$~keV ($E_5$ in Table~\ref{tab:snrfits}) is 144.5~eV. Additionally, as mentioned in Section~\ref{subsec:fano}, the instrument noise performance was also verified on-board by acquiring data in the said mode, and the average noise values obtained ($\sim 10~e^-$~rms) were similar to those obtained on ground.

\subsection{Payload alignment determination}
\label{subsec:align}
As mentioned in Section~\ref{subsec:xsp}, \xsp has square collimators, giving it a triangular response parallel to its sides (see Figure~4 in [\citenum{rkrish2025}]). The spacecraft bus ensures the POLIX view axis is kept within $\pm 0.1^{\circ}$ of the source direction. The two \xsp detector packages, in turn, are aligned with respect to POLIX axis on ground. However, manufacturing errors (e.g. flatness) and mounting inaccuracies can lead to slight deviations from the expected alignment. Moreover, the alignments can further change due to launch stresses and vibrations. Hence we carried out scan observations along the two orthogonal axes to determine the pointing axis of each collimator. Crab was chosen for this operation because it is a bright, steady source with a known flux. 

\begin{figure*}
\centering
\includegraphics[scale=0.64, trim=1cm 2cm 0cm 2cm, clip=true]{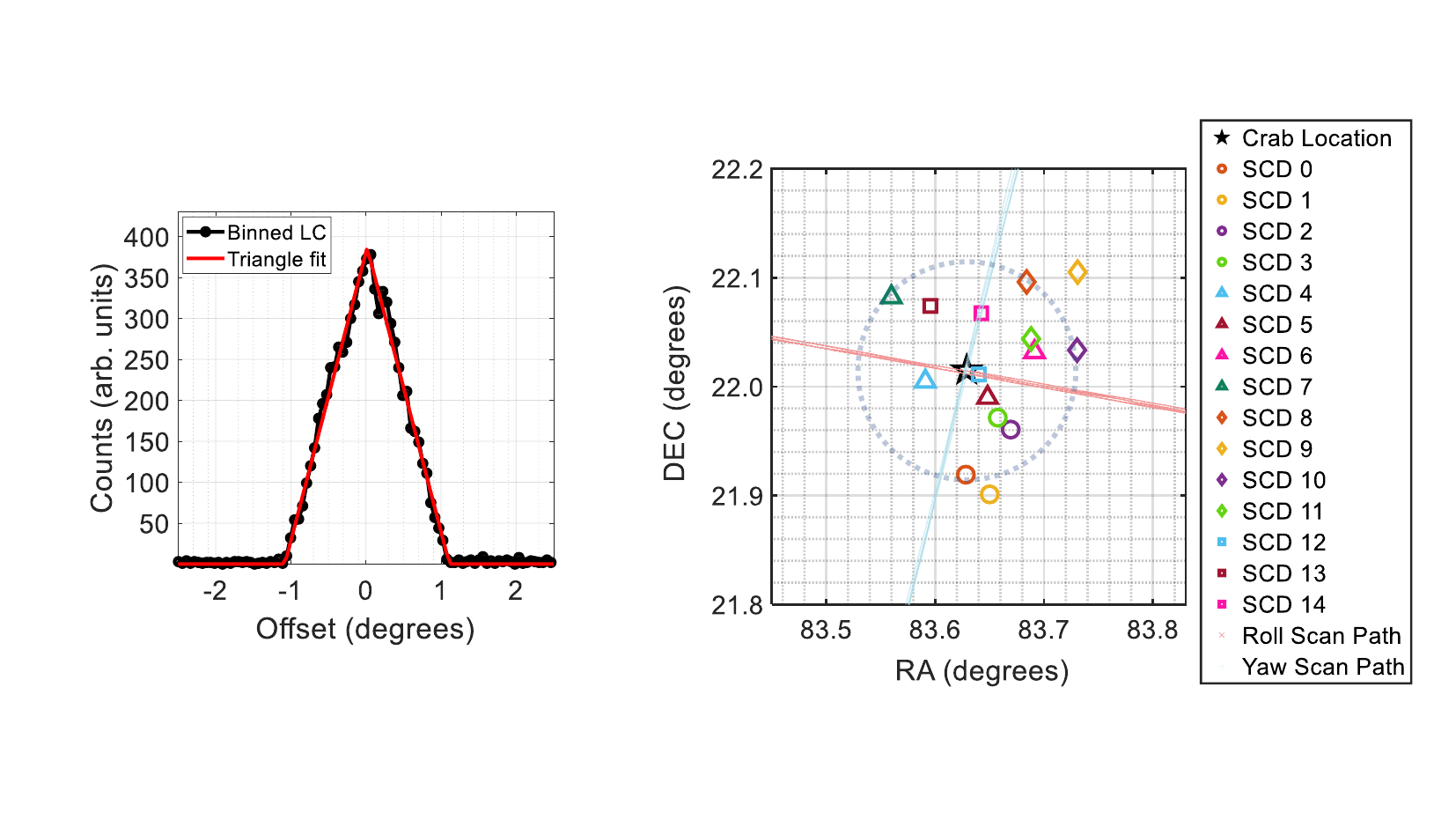}
\caption{Collimator offset determination using Crab scan observations: (a) Fit to binned lightcurve using a triangular profile. (b) Determined offsets from the pointing direction for each collimator. The dotted circle has a radius of $0.1^{\circ}$}
\label{fig:collalign}
\end{figure*}

The scans were performed such that the scan path crosses the source location. In each orbit, the spacecraft was scanned $\sim \pm 6.5^{\circ}$ across the source at a constant rate (0.0058 degrees/s). To determine the pointing direction of each collimator, the observed flux across the scan was binned as a function of offset angle and fitted with a symmetric triangular profile (Figure~\ref{fig:collalign}a). The angular distance of the vertex from the Crab location gives the collimator offset. Figure~\ref{fig:collalign}b shows the measured offsets along these two axes ($\theta_{x,i}, \theta_{y,i}$), for all fifteen open detectors. Based on these measurements and the known collimator response function, the flux correction factors have been determined and the \xsp effective area has been updated (Section~\ref{sec:arf}). 

\subsection{Fine-tuning of effective area}
\label{sec:arf}
The instrument response consists of the spectral redistribution function as well as the effective area. The former was determined experimentally using ground experiments (Section~\ref{subsec:rrcat}). The latter includes the following multiplicative components for the case of \xsp:
\begin{enumerate}[label=(\Alph*)]
\item Device geometrical area
\item Correction factor due to collimator alignment
\item Collimator open area fraction
\item Optical light filter transmission
\item Detector quantum efficiency (QE)
\item Single event correction factor
\end{enumerate}

Of these, A - C are purely geometrical factors, whereas D - F have energy-dependence as well. The device active area (factor~A) is $\sim 4$~cm$^2$ [\citenum{holland2008}].

\begin{figure*}
\centering
\includegraphics[scale=0.64]{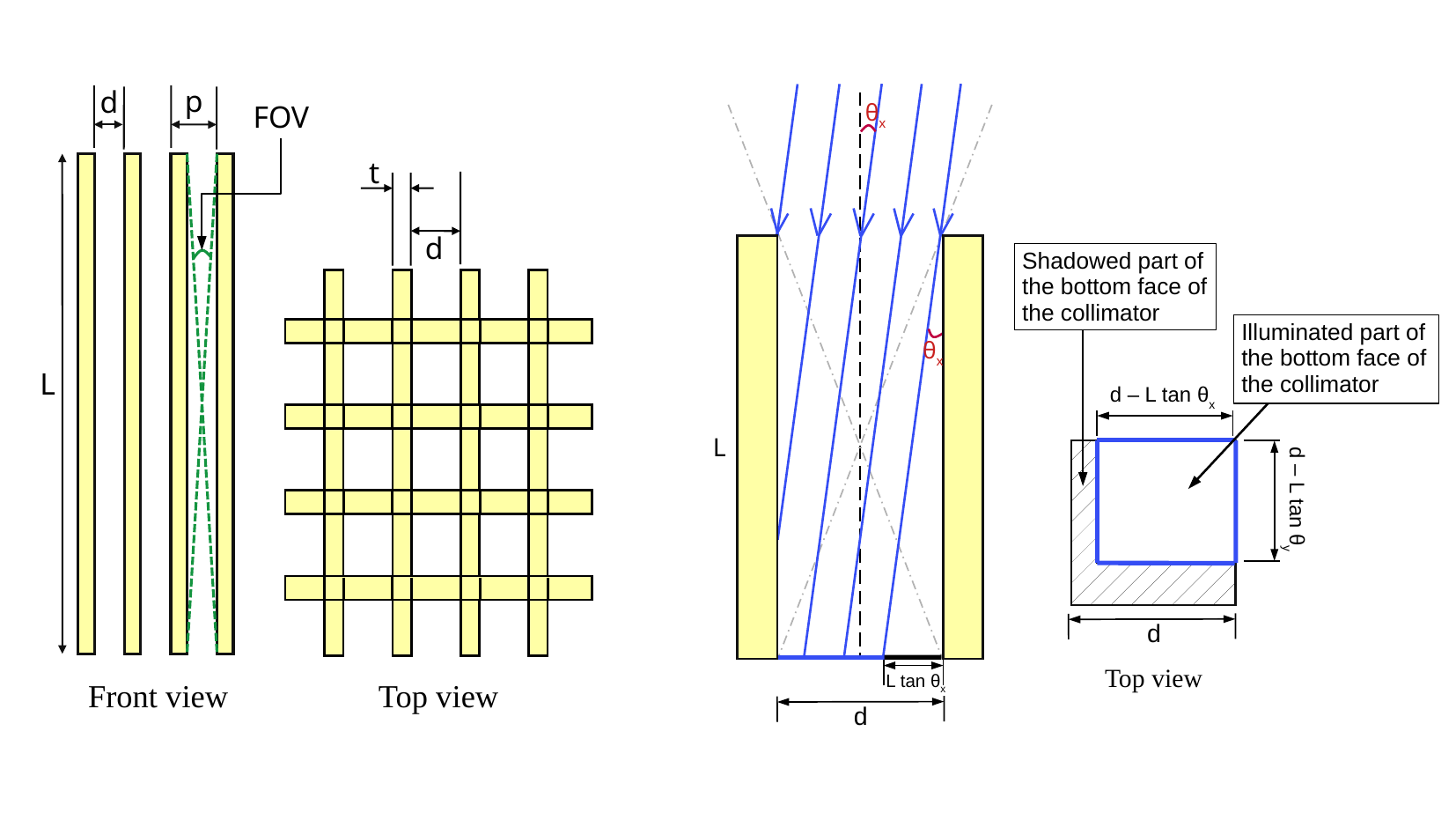}
\caption{Schematic (not to scale) of (a) the collimator structure, depicting the length (L), cell size (d), wall thickness (t), pitch (p) and field of view (FOV), and (b) illustration of a source observation with an offset $\theta_x$ from the collimator axis. An exaggerated view of a single collimator cell is also shown, demonstrating the ACF calculation. See text for details.}
\label{fig:colli}
\end{figure*}

The individual collimator alignments have been measured on-board (Section~\ref{subsec:align}). However, the corresponding alignment correction factor (ACF; factor~B) to the effective area depends on the FOV, which in turn depends on the wall thickness ($t$) of the collimators. The parameter $t$ also affects the collimator open area fraction (OAF; factor~C). The ACF, FOV and OAF are all dependent on $t$ as follows:
\begin{equation}
FOV_j (t_j) = \left[2\tan^{-1}\left(\frac{p_j-t_j}{L}\right)\right]^2
\end{equation}
\begin{equation}
OAF_j (t_j) = \left(\frac{p_j-t_j}{p_j}\right)^2
\end{equation}
where $j=2,3$ indicates the type of collimator, $p_j$ is the pitch of each collimator cell, $t_j$ is the wall thickness and $L$ is the height of the collimator, as illustrated in Figure~\ref{fig:colli}a. The $ACF$ for each device can be derived as a function of the alignment angle (distance from perfectly on-axis alignment) by calculating the shadowed fraction of the top face of the collimator on the bottom face, assuming a `misaligned' source at infinity (Figure~\ref{fig:colli}b). Defining the open part of each cell $d_j = p_j - t_j$, 
\begin{equation}
ACF_i (t_j, \theta_{x,i}, \theta_{y,i}) = \frac{d_j^2 - d_jk_{x,i} - d_ik_{y,i} + k_{x,i}k_{y,i}}{d_j^2} 
\end{equation}
where $k_{x,i}=|L\tan\theta_{x,i}|$, $k_{y,i}=|L\tan\theta_{y,i}|$, and $\theta_{x,i}$, $\theta_{y,i}$ are the measured collimator alignments ($i = 0-14$).

The collimator height $L$, and pitch $p_2$ ($=d_2+t_2$) and $p_3$ ($=d_3+t_3$) are accurately measured and well-constrained. However, fabrication tolerances can lead to different `effective' values of $t_2$ and $t_3$. The $3^{\circ}\times 3^{\circ}$ collimators, manufactured by EDM wire-cutting, have sharper, well-defined inner wall edges which enables an accurate measurement of the wall thickness on a profile projector. In contrast, due to manufacturing limits, the smaller FOV could not be achieved using wire-cutting technique, and they were fabricated by additive manufacturing. Hence, the $2^{\circ}\times 2^{\circ}$ collimators have coarser wall edges which are difficult to be measured on a profile projector, therefore we proceed to fine-tune this value using on-board observations as explained below.

\begin{figure}[h!]
\centering
\includegraphics[scale=0.6, trim=3cm 2cm 3cm 3cm, clip=true]{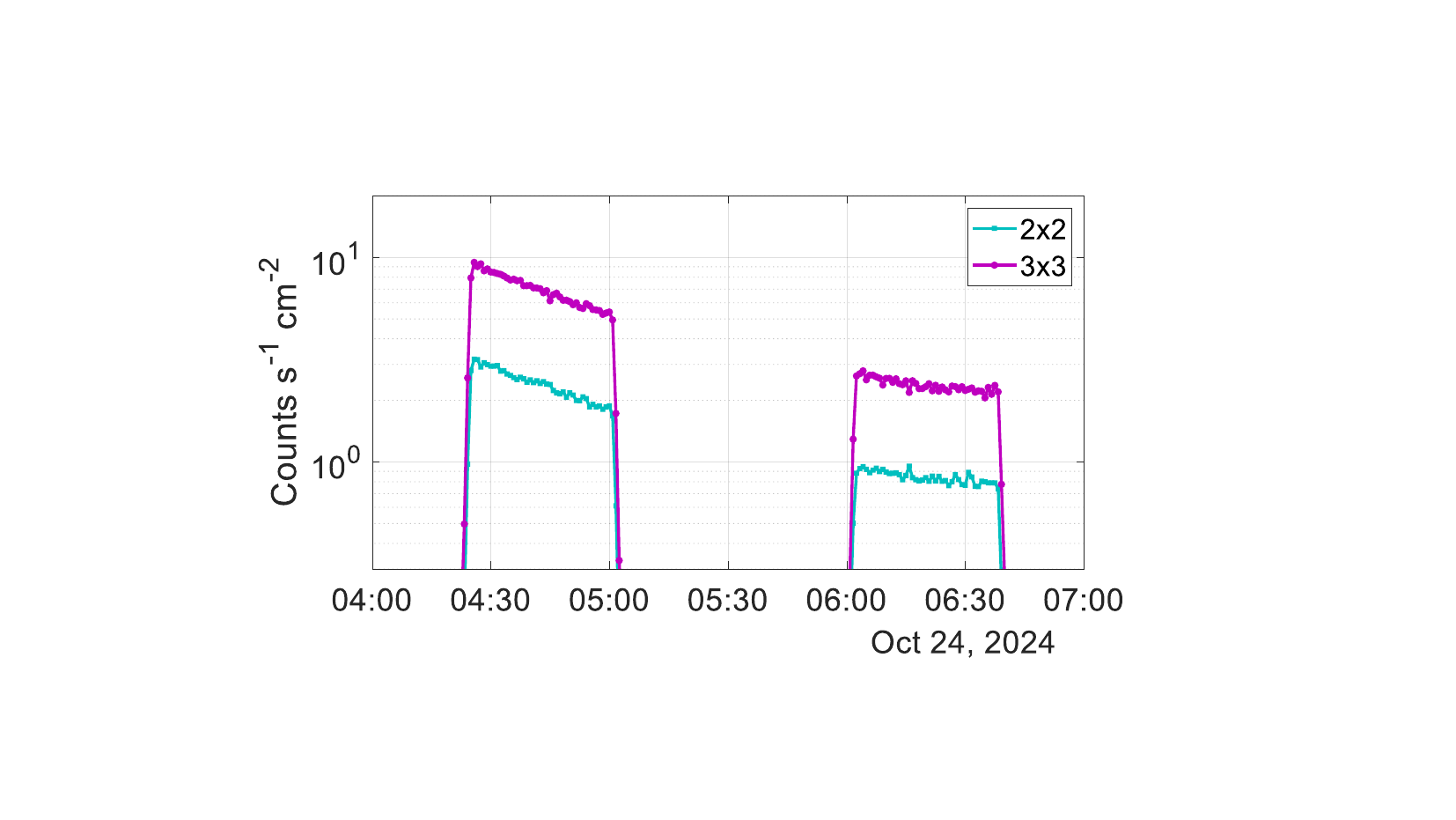}
\caption{Day-side light curve of \xsp during a Solar flare, depicting the ratio of count rates detected by the two kinds of detectors. }
\label{fig:dayratio}
\end{figure} 

The observed counts from the two different kinds of detector vary only in terms of collimator-dependent factors, viz. ACF, OAF and FOV, all of which can be parameterized by the two variables $t_2$ and $t_3$ (equations $4-6$). Hence, for a source which uniformly illuminates the entire field of view of all the detectors, the ratio of observed counts ($R$) by two different kinds of detector can be used to indirectly infer the value of $t_2$ (having known measurement of $t_3$) as:
\begin{equation}
R = \frac{\mathrm{Counts_3}}{\mathrm{Counts_2}}=\frac{\mathrm{OAF_3}}{\mathrm{OAF_2}}\times\frac{\mathrm{FOV_3}}{\mathrm{FOV_2}}\times \frac{\langle ACF\rangle_{\mathrm{3\times 3}}}{\langle ACF\rangle_{\mathrm{2\times 2}}}
\end{equation}
where $\langle ACF\rangle_{\mathrm{j\times j}}$ denotes the average over the respective field of view detectors. Blank sky observations (Section~\ref{sec:bkg}) provide an opportunity to carry out this exercise. However, statistical variations due to the low counts in the X-ray background make the ratio (R) determination prone to errors. A better and more robust method is to utilise the day side observations, when \xsp is looking at the Earth (Figure~\ref{fig:dayratio}) and the devices are uniformly illuminated. During Solar flares, the reflected radiation from the upper layers of the Earth's atmosphere has significant counts in \xsp. With a measured average $R=2.9$ for multiple such flares, we obtain $t_2$. Finally, using this estimated value of $t_2$ and the measured value of $t_3$, the actual FOVs ($1.95^{\circ}\times 1.95^{\circ}$ and $2.89^{\circ}\times 2.89^{\circ}$), FOV-averaged ACFs (88.2\% and 94.5\%) and OAFs (59.5\% and 73.5\%) have been computed for $2^{\circ}\times 2^{\circ}$ and $3^{\circ}\times 3^{\circ}$ detectors respectively.

\begin{figure*}
\centering
\includegraphics[scale=0.6, trim=0cm 3cm 0cm 3cm, clip=true]{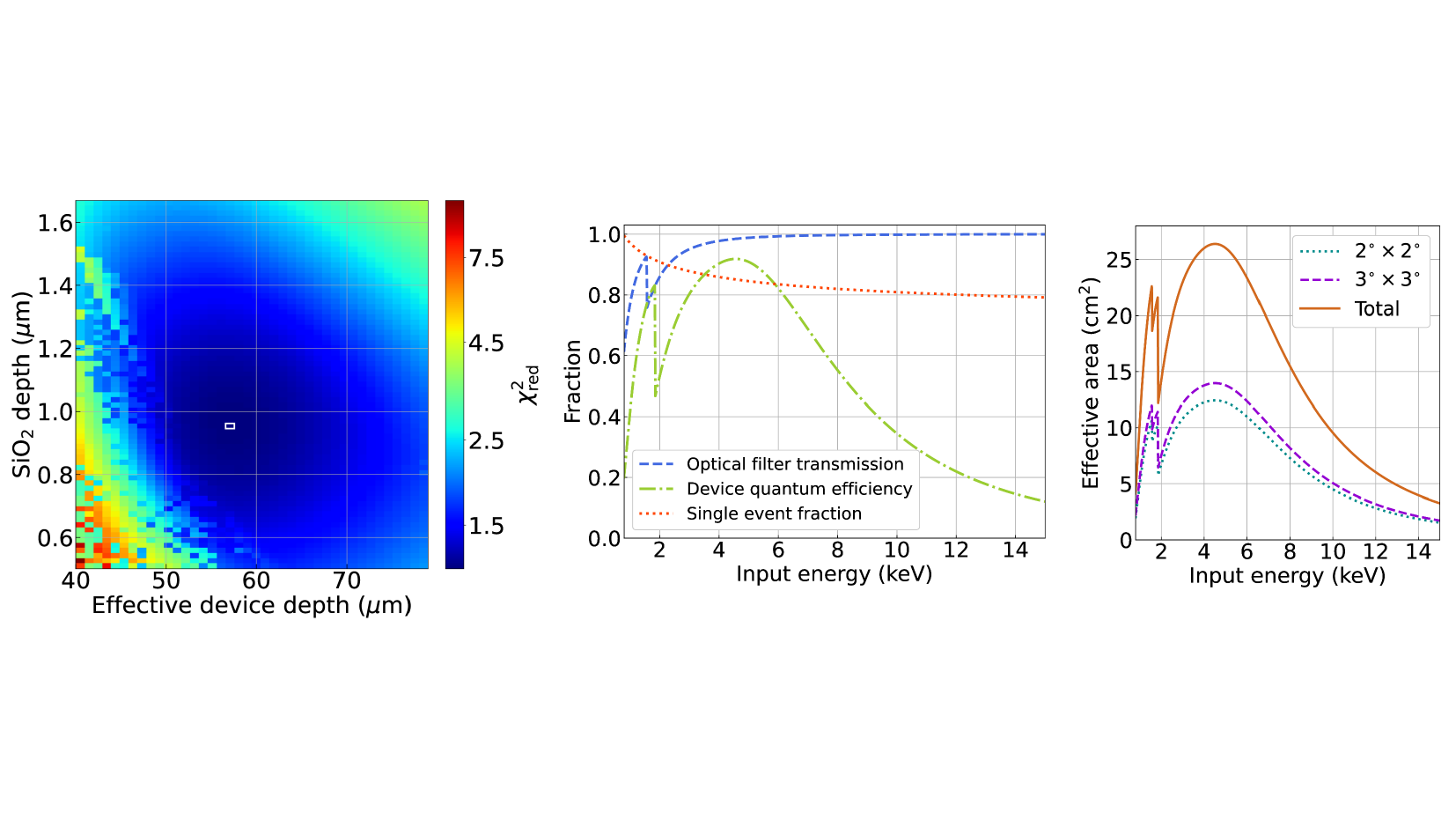}
\caption{(a) Results of iterative fitting of Crab spectra using different instrument effective areas. QE has been defined over a grid of device depth and dead layer thicknesses. The $\chi^2_{\mathrm{red}}$ value is shown as color, and the white pixel marks the best fit. (b) Energy-dependent factors in the effective area. The QE shown corresponds to the best fit in (a). (c) Net single event effective area of \xsp ($0.8-15$~keV). The contribution of the detectors with two kinds of collimators is also shown.}
\label{fig:arftune}
\end{figure*} 

Further, among the energy-dependent factors, factor~D is well-constrained due to the known thicknesses of the different layers of the filter, and factor~F is experimentally obtained using ground experiments (Section~\ref{sec:sinevt}). As for the device QE (factor~E), even though the approximate layer thicknesses are known [\citenum{smith2014,athiray2015}], the knowledge of the total \textit{effective} depth from which charge collection, both total as well as partial, takes place is not known. Moreover, at the synchrotron facility where SRF determination experiments were carried out (Section~\ref{subsec:rrcat}), the beam flux was variable during the course of the experiments, preventing an absolute QE measurement. Hence, we used on-board Crab observations to determine the QE. We generated effective areas over a grid of parameters, by varying the effective device depth as well as the top (dead) layer thickness, and fitted Crab observations with a canonical absorbed power law model. Figure~\ref{fig:arftune}a shows the reduced chi-squared ($\chi^2_{\mathrm{red}}$) values of the spectral fits. The parameters which produce the best fits (lowest $\chi^2_{\mathrm{red}}$) are 57~$\mu$m and 952~nm respectively for the effective depth and top layer thickness, and these values are used to generate the QE (factor~E). The factors D - F are shown in Figure~\ref{fig:arftune}b, and the net effective area of the $2^{\circ}\times 2^{\circ}$ and $3^{\circ}\times 3^{\circ}$ detectors (eight and seven SCDs respectively) are shown in Figure~\ref{fig:arftune}c.

\subsection{Response validation: Crab spectral fit}
\label{sec:crabfit}

\begin{figure*}
\centering
\includegraphics[scale=0.68, trim=4cm 0cm 4cm 0cm, clip=true]{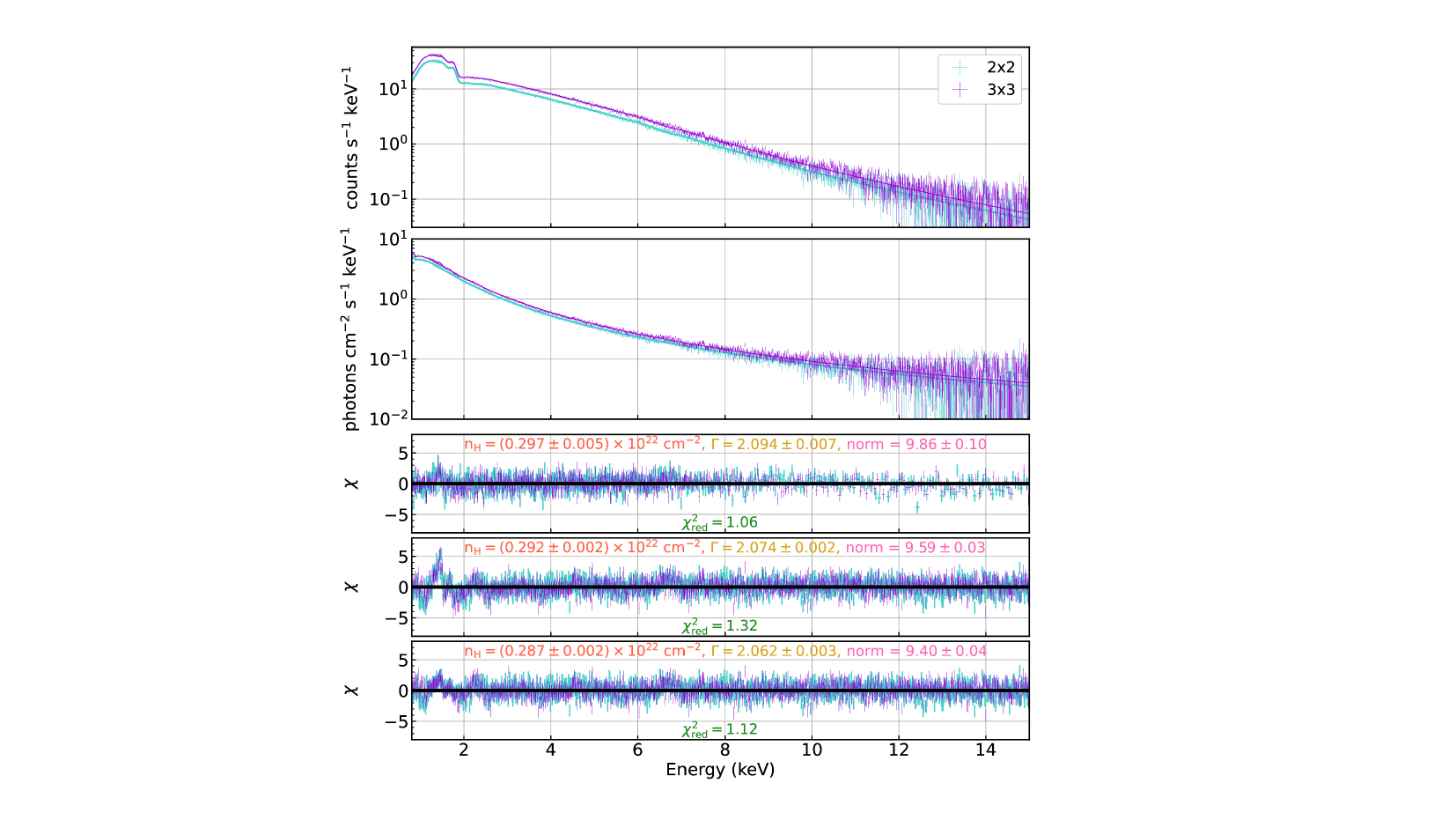}
\caption{Crab spectral fit using an absorbed powerlaw model. The bottom three panels show the residuals (in units of $\sigma$) on fitting 1 orbit data, 1 day data without systematics, and 1 day data with 3\% systematics respectively. Best fit parameters are mentioned in the respective panels. The top two panels respectively depict the observed and unfolded spectra corresponding to the fifth panel.}
\label{fig:crabfit}
\end{figure*}

The instrument response includes the ARF (ancilliary response file), which describes the effective area of the instrument, and the RMF (redistribution matrix file), describing the device spectral response, i.e. SRF. We fitted the background-subtracted Crab spectrum for a validation of these response files. First, we extracted the spectrum of the two FOVs for a single orbit of observations (T$_{\mathrm{exp}} \sim 1.4$~ks). The spectra were simultaneously fitted with a canonical absorbed powerlaw model (\texttt{tbabs*pow} in \textit{XSPEC} [\citenum{arnaud1996}]), as typically used for Crab [e.g. \citenum{kirsch2005} and references therein]. A constant multiplicative factor was included to account for the cross-calibration between the two FOVs, which was frozen to 1 for $2^{\circ}\times 2^{\circ}$ and left free to vary for $3^{\circ}\times 3^{\circ}$. All other parameters (hydrogen column density N$_{\mathrm{H}}$, powerlaw index $\Gamma$, and powerlaw norm) were tied across the two spectra. We obtained a good fit ($\chi^2_{\mathrm{red}}=1.06$) with this model, with parameter values and residuals shown in the third panel of Figure~\ref{fig:crabfit}.

Next, we take the entire day's data (T$_{\mathrm{exp}} \sim 12.0$~ks) and fit it with the same model. The residuals and fit parameters are shown in the fourth panel of Figure~\ref{fig:crabfit}. We find systematic residuals in the energy range $\sim 1.2-1.8$~keV, resembling a Gaussian-like feature. This is likely an instrumental feature due partly to fluorescence from the aluminized optical light blocking filter, excited by the source photons and getting absorbed into the active volume. It is to be noted that this feature is prominent only for very bright sources ($\gtrsim (1-2)\times 10^6$~counts~per~FOV). For longer integrations (or higher spectral counts), the inclusion of two \texttt{edge} components, corresponding to the K-absorption edges of Al and Si, may provide a better fit at low energies ($< 2$~keV). Efforts are currently underway to model and include these effects as part of the instrument response. 

Even though including an additional Gaussian line in the model eliminates the residual $\sim 1.5$~keV, we do not recommend this as the underlying source spectral parameters may be affected by the line parameters. Instead, we suggest adding upto $\sim 3\%$ systematics to the spectra in this energy range while fitting the spectra of bright sources. Note that this percentage may have to be adjusted depending on the flux of the source. The spectral fit after including systematics is shown for the full day Crab data in panels 1, 2 and 5 of Figure~\ref{fig:crabfit}. The fit parameters are consistent with those found using other instruments [\citenum{kirsch2005}], and hence we conclude that the response files adequately capture the instrument spectral behaviour. 

\subsection{Background}
\label{sec:bkg}

Every space-based instrument is subjected to incidence of various sources of radiation. The radiation sources vary depending on the orbit of the spacecraft, the attitude of the telescope/instrument, the energy band in which the observations are made, etc. The interaction of these sources with the instrument on-board has profound effects on the overall performance of the payload and also the net output signal from it. For \xsp, which is a collimated soft-X-ray spectrometer in a low Earth orbit, the source signal from an astronomical object gets entwined with signal due to the following sources, thereby altering the resultant spectrum from the instrument:

\begin{enumerate}[label=(\alph*)]
\item Cosmic X-ray background (CXB): CXB [e.g. \citenum{abdo2010}] consists of Galactic and extra-Galactic sources (both resolvable and un-resolvable) that are present in the FOV of the instrument, which interact directly with the detector to generate background events. 
\item Galactic cosmic rays (GCRs): GCRs [\citenum{simpson1983, mewaldt1994}] result from the extremely high-energy astrophysical processes wherein protons, heavy ions, and electrons with energy spanning from few MeVs to TeVs and beyond get generated. Owing to the high energy of these particles, they may pass through any shielding and reach the SCDs for primary interaction, as well as generate secondary and tertiary particles/photons that may eventually hit the SCDs simultaneously. 
\item Trapped energetic particles in the SAA: SAA is a regional phenomenon where enhanced proton and electron flux is prevalent in a region above the south-Atlantic ocean [\citenum{dessler1959, heirtzler2002}]. Although \xsp is powered off during SAA passage, the `SAA' definition used on-board XPoSat is spatially static and hence, energetic particles may still enter the detectors owing to the dynamic nature of Earth's geomagnetic bubble [e.g. \citenum{li2025}].
\item  Transients: Solar transients such as solar energetic particles (SEPs) and coronal mass ejections (CMEs) can enter Earth's magnetosphere to reach the location of the satellite and add to the background. Apart from solar activity-induced events, there can also be magnetospheric precipitation events [\citenum{liao2020}] which may show up as background counts. 
\end{enumerate}

\begin{figure*}
\centering
\includegraphics[scale=0.62, trim=0cm 5cm 0cm 5cm, clip=true]{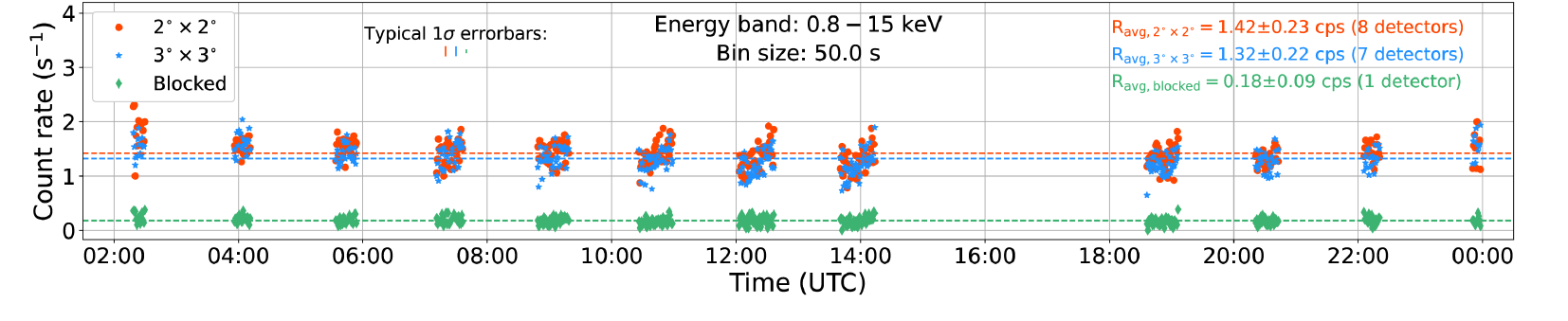}
\caption{Typical blank sky lightcurve of \xsp (2024~April~25). The day-averaged count rates of $2^{\circ}\times 2^{\circ}$, $3^{\circ}\times 3^{\circ}$, and blocked detectors are marked with red, blue and green dashed lines respectively.}
\label{fig:bkglc}
\end{figure*}

To estimate the factors (a) and (b), we carried out several pointings of blank sky regions having no known X-ray sources with flux higher than the instrument sensitivity, with a net accumulated exposure of $> 1$~Ms. The $0.8-15$~keV light curve of a typical blank sky observation is shown in Figure~\ref{fig:bkglc}. From the average background count rate across all the blank sky observations, the $5\sigma$ sensitivity of \xsp is estimated to be $\sim 0.6$~mCrab in 10~ks.

\begin{figure*}
\centering
\includegraphics[scale=0.6, trim=4cm 3cm 4cm 3cm, clip=true]{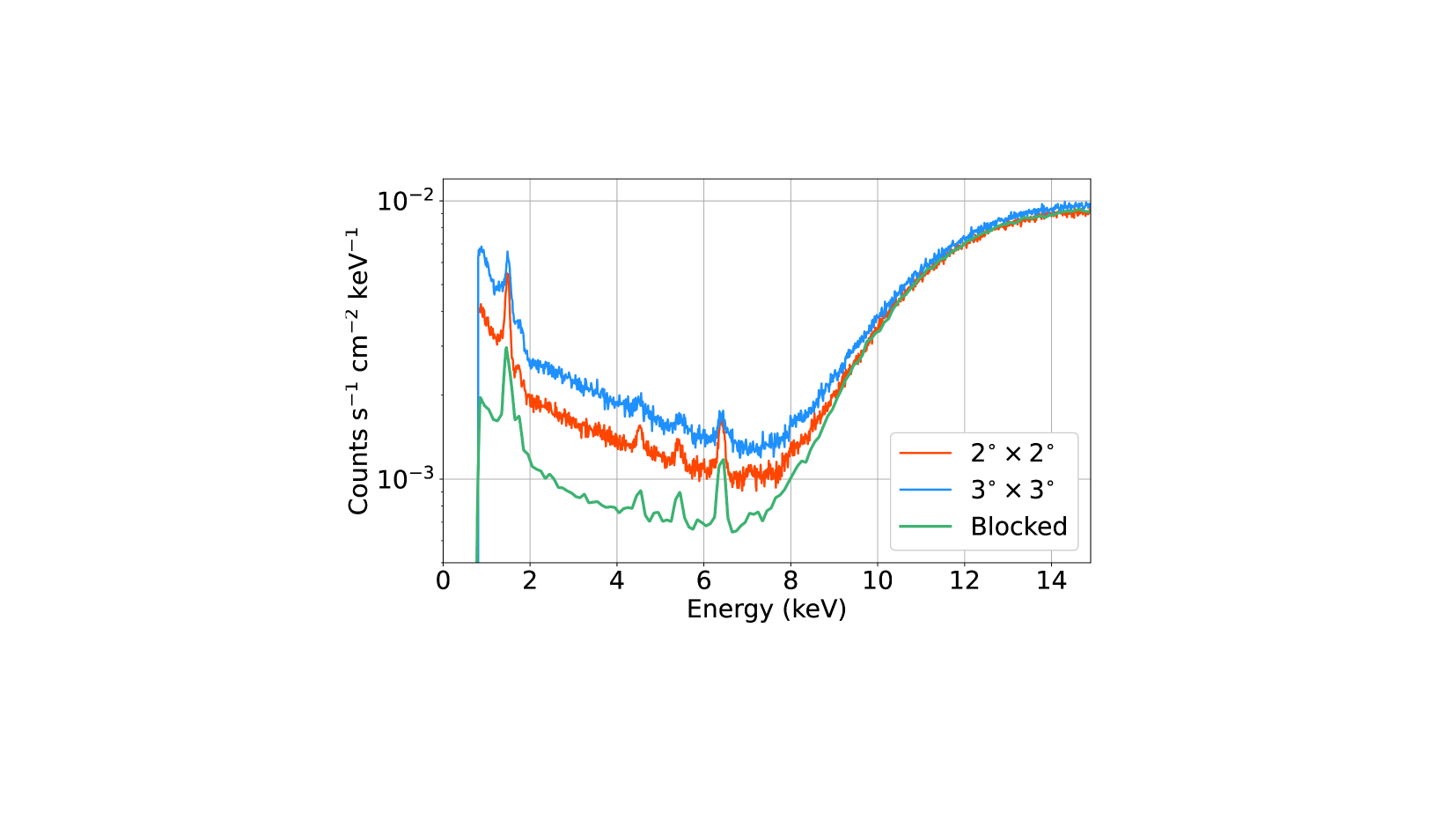}
\caption{Typical blank sky spectrum of \xsp, observed by the $2^{\circ}\times 2^{\circ}$, $3^{\circ}\times 3^{\circ}$, and blocked detectors.}
\label{fig:bkgspec}
\end{figure*} 

A detailed analysis and systematic modeling of the background observations is ongoing, and will be presented in a separate communication. However, to first order, the background count rates are comparable across the different pointings. The average background spectra observed by the two FOVs is shown in Figure~\ref{fig:bkgspec}. The spectrum of the blocked detector, which is mainly sensitive to the GCRs, is also shown. CXB photons, entering through the collimators, usually dominate in the lower energies ($\lesssim 8$~keV) whereas the GCRs, impinging from all directions, dominate at higher energies and has a similar contribution for all detectors (the Tantalum piece over the blocked detector produces fluorescence in the \xsp energy band, which can be modeled out). 

For spectral analysis, the averaged blank sky spectrum gives a good estimate of the background, and can be directly used (`static' background). Further, it has been observed over the past 1 year of observation that the blocked detector spectral shape, mainly due to GCRs, has remained steady, whereas its absolute scale shows gradual variation with solar cycle [\citenum{zeitlin2019}]. Moreover, considering the entire $0.8 - 15$~keV energy band, the GCR contribution dominates over that of CXB (GCR $\sim 2.6$~cps vs CXB $\sim 0.3$~cps for 15 detectors). To capture this variation, a provision is made in the \xsp software to estimate the background during a particular source observation, capturing the level of particle background at that time (`hybrid' background). In this mode, the CXB is assumed to be static, but the GCR component is scaled for the specific observation using the saturated count rate (explained below). 

\begin{figure*}
\centering
\includegraphics[scale=0.6, trim=0cm 4cm 0cm 2cm, clip=true]{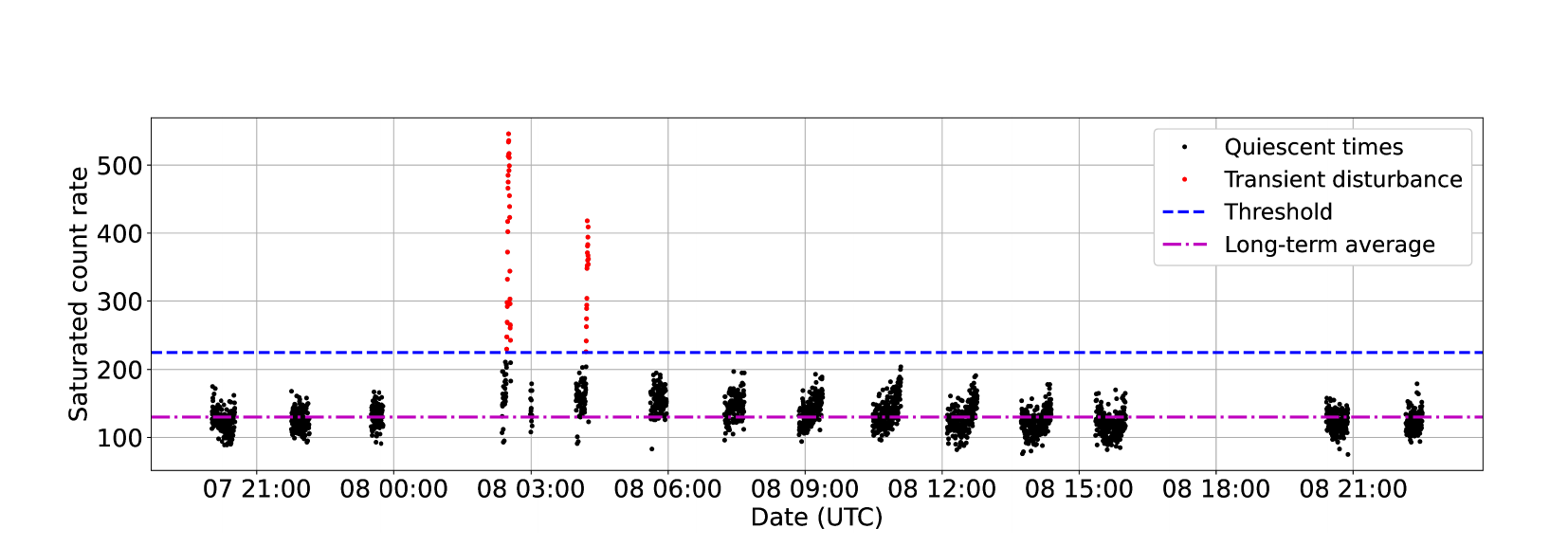}
\caption{Light curve of saturated counts during quiescent and disturbed times (2024~June~$7-8$), along with threshold for filtering. }
\label{fig:satlc}
\end{figure*}

In addition to the regular background, there may be transient events not attributed to the astrophysical source being observed (factors (c) and (d)). Primarily, these events are observed near SAA entry or exit, and show up as temporary enhancements throughout the energy band, mimicking X-ray events. Any event which deposits energy higher than 15~keV is recorded as counts in the saturation channel. The saturation count rate is continuously monitored, and is found to be extremely stable in the long-term (Figure~\ref{fig:satlc}). Any spikes (defined as $> 4\sigma$ above the long-term average quiescent rate) in this count rate is a telltale sign of particle events, and provision is made in the \xsp software to identify and remove such durations. 

Note that as mentioned in Section~\ref{subsec:xsp}, the presence of two different FOVs along with a blocked detector make ``in-situ'' measurement (during the source observation itself) of the background  possible. We are currently evaluating the feasibility of using this method for a better estimate of background.
 
\subsection{Verification of timing capabilities}
\label{sec:timing}

Some of the science objectives of \xsp require carrying out timing and phase-resolved spectroscopy of pulsars. Hence, it is important to first establish the timing capabilities of the instrument. Crab is a rotation-powered pulsar which shows modulations at $\sim 33$~ms. Since the timing properties of Crab are well-characterised, we used observations of this source to verify the timing capabilities of \xsp.

\begin{figure*}
\centering
\includegraphics[scale=0.6, trim=0cm 4cm 0cm 2cm, clip=true]{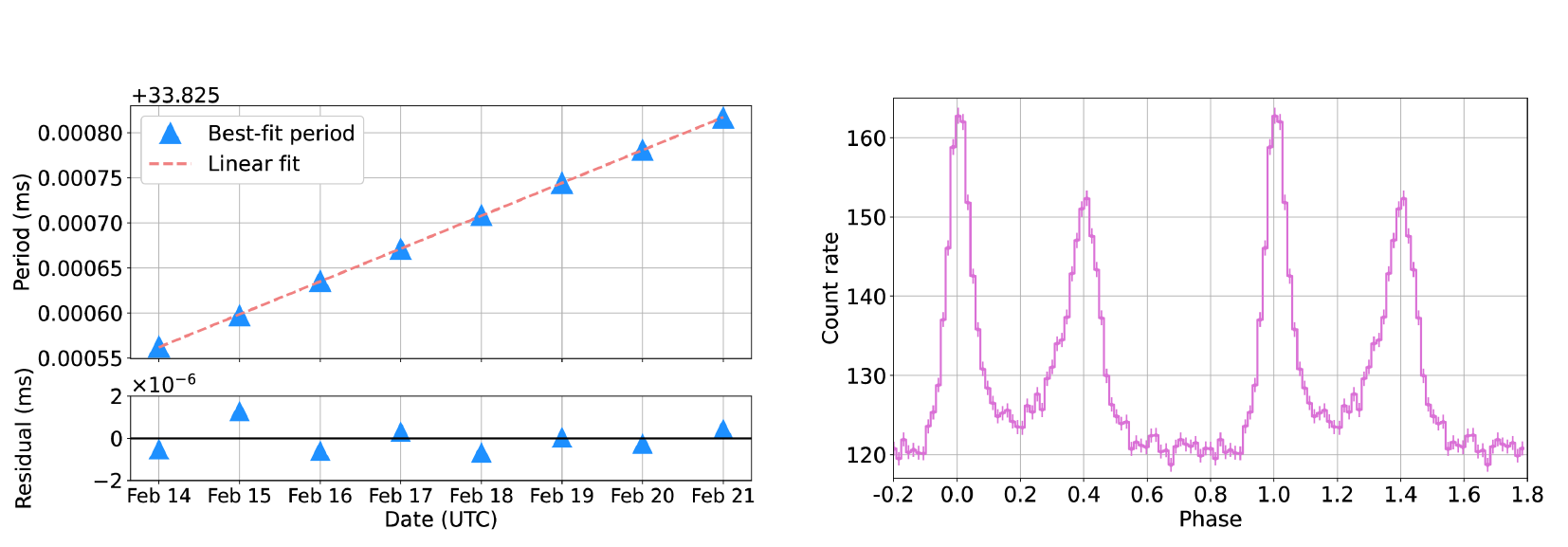}
\caption{(a) Best-fit period of Crab, determined day-wise, in the duration 2024~February~14 to 18. A linear fit to the periods is overlaid. (b) Pulse profile of Crab in $0.8-10$~keV, depicting the typical two-horned structure.}
\label{fig:pdot}
\end{figure*} 

For this exercise, we used Crab observations corresponding to $14-21$~February~2024. The event arrival times in the screened (L2) event list were first corrected to correspond to the solar system barycenter. Day-wise light curves in $0.8-10$~keV with 1~ms binning were generated from the corrected event file. Taking the Jodrell Bank radio ephemeris of Crab as the starting point [\citenum{lyne1993, jbcrab}], we searched for pulsations in \xsp data using \textit{efsearch} [\citenum{efsearch}], which employs the epoch-folding method to detect periodicities in a time series [\citenum{leahy1983}]. Strong pulsations were detected in all the datasets, increasing from 
0.03382556260(9)~s on 2024~February~14 to 0.03382581678(5)~s on 2024~February~21. To estimate the uncertainties, 500 light curves were simulated corresponding to each observation, and the dispersion in the periods determined from the simulated light curves are used [\citenum{boldin2013}]. The average difference between the periods determined from \xsp data and those computed from Jodrell Bank ephemeris is $\sim 0.59~\mu s$. We determined the period derivative over this eight day period to be $\dot{P}\sim 4.22\times 10^{-13}$~s~s$^{-1}$ by fitting a linear function to the obtained best periods (Figure~\ref{fig:pdot}a).

The typical pulse profile in the $0.8-10$~keV energy band, obtained by folding the light curve over the determined best-fit period, is shown in Figure~\ref{fig:pdot}b. The pulse fractions of the main and secondary peaks in the $0.8-10$~keV band are 15.1\% and 11.9\% respectively. Further, we extracted light curves in four energy bands, viz. $0.8-2$~keV, $2-4$~keV, $4-6$~keV and $6-10$~keV, and generated the corresponding pulse profiles. For both main and secondary peaks, the pulse fraction is found to increase with energy ($14.2 - 20.8$\% and $10.9 - 17.0$\% respectively).  

\section{Summary}
In this paper, we have described both on-ground and in-flight calibration of \xsp instrument on-board XPoSat in detail. The ground tests and calibration included thermo-vacuum qualification, followed by the characterization of temperature-dependent gain and resolution of each device, across the entire range of operating temperatures. The spectral redistribution function across the \xsp energy band was constructed using data from monochromatic beamline experiments. The noise performance of the instrument was also characterised. 

Following launch, a series of on-board observations were carried out during the Performance Verification phase to refine and validate these calibrations. These included observations of standard calibration sources such as Cassiopeia A, Tycho, and the Crab pulsar. The gain and resolution parameters derived from ground tests were confirmed through line energy and width measurements. Individual collimator alignment offsets were measured using spacecraft scan observations of the Crab, enabling angular corrections to the effective area. Being a bright and standard source, Crab observations were also used to fine-tune the \xsp effective area. The background was characterized using long-term blank-sky observations and monitoring of particle fluxes, enabling both static and dynamic background estimation schemes. Additionally, timing capabilities were verified through pulse period analysis of the Crab pulsar, establishing XSPECT’s suitability for millisecond timing studies.

Together, the results from both ground tests and in-flight observations demonstrate that \xsp is well-calibrated and establishes its performance for spectro-temporal studies of astrophysical sources in the soft X-ray band. Even though there are contemporary instruments with higher sensitivities, the uniqueness of \xsp lies in the continuous long-term monitoring of sources, which can provide a wealth of valuable data and crucial insights in this field. Moreover, as compared to imaging instruments, \xsp provides much better time resolution, and overcomes the problem of pile-up -- making it a suitable instrument to study bright transients as and when they are discovered. \xsp has now been opened up for proposal-based community-driven observations [\citenum{xpps}], similar to AstroSat. All necessary software tools, user guides, as well as response and background files are available in the public domain to be utilized by the scientific community. The data hosted at ISSDC can be accessed via PRADAN [\citenum{pradan}].

\subsection*{Disclosures}
The authors declare that there are no financial interests, commercial affiliations, or other potential conflicts of interest that could have influenced the objectivity of this research or the writing of this paper.

\subsection*{Code, Data, and Materials Availability} 
The data utilized in this study from ground experiments and calibration are available from the authors upon request. The source observation data used are also presently available from authors upon request while they are being made available at PRADAN. \textit{NICER} data of Cas~A has been obtained from HEASARC data archive (https://heasarc.gsfc.nasa.gov/cgi-bin/W3Browse/w3browse.pl).  

\subsection*{Acknowledgments}
We thank the XPoSat project team, facilities team, assembly, integration, and checkout teams, and mission team for their involvement and support in enabling XSPECT payload on XPoSat mission. We thank Director, U R Rao Satellite Centre, Deputy Director, PDMSA, and Group Head, SAG for their reviews and support. We would like to extend our gratitude to Dr. Mohammed H. Modi, Dr. M.K. Tiwari, Dr. S. K. Rai and their teams for facilitating and supporting our experiments at the Indus beamlines, RRCAT.

\bibliography{references}   
\bibliographystyle{spiejour}   

\vspace{1ex}
\noindent Biographies of the authors are not available.

\end{spacing}
\end{document}